\documentclass[preprint,amsmath,amssymb,nofootinbib]{revtex4}
\usepackage[dvips]{graphics}
\usepackage{graphics}
\makeatletter
\renewcommand{\fnum@figure}{Fig. \thefigure}
\makeatother
\usepackage{pstcol}
\usepackage{amsfonts}
\usepackage{bm}
\usepackage{amsmath}
\usepackage{amssymb}
\usepackage{color}
\usepackage[all]{xy}
\usepackage{comment}
\usepackage{epstopdf,epsfig}
\def\be{\begin{equation}}
\def\ee{\end{equation}}
\def\bea{\begin{eqnarray}}
\def\eea{\end{eqnarray}}
\usepackage{float}

\begin{document}

\title{Phase transitions, shadows, and microstructure  of Reissner-Nordstr\"om-Anti-de-Sitter black holes from a geometrothermodynamic perspective}

\author{Jose M. Ladino$^1$, Carlos E. Romero-Figueroa$^1$, and Hernando Quevedo$^{1,2,3}$} 
\email{miguel.ladino@correo.nucleares.unam.mx;carlosed.romero@correo.nucleares.unam.mx;quevedo@nucleares.unam.mx}

\affiliation{$^1$Instituto de Ciencias Nucleares, Universidad Nacional Aut\'onoma de M\'exico, AP 70543, Mexico City, Mexico}
\affiliation{$^2$Dipartimento di Fisica and Icra, Universit\`a di Roma “La Sapienza”, Roma, Italy.}
\affiliation{$^3$Al-Farabi Kazakh National University, Al-Farabi av. 71, 050040 Almaty, Kazakhstan.}

\date{\today}

\begin{abstract}

We study the thermodynamic properties of the Reissner-Nordstr\"om black hole with cosmological constant, expressed in terms of the curvature radius, by using the approach of shadow thermodynamics and the formalism of geometrothermodynamics. We derive  explicit expressions for the shadow radius in terms of the horizon, photon sphere, and observer radii. The phase transition structure turns out to strongly depend on the value of the curvature radius, including configurations with zero, one, or two phase transitions. We also analyze the black hole microscopic structure and find differences between the approaches of thermodynamic geometry and geometrothermodynamics, which are due to the presence of the curvature radius. We impose the important condition that the black hole is a quasi-homogeneous thermodynamic system to guarantee the consistency of the geometrothermodynamic approach.


\end{abstract}


\maketitle

\section{Introduction}
\label{sec:int}

Recent outstanding observational achievements, including the detection of gravitational waves and the observation of supermassive black holes, have increased the interest in investigating the properties of black holes in Einstein theory and its generalizations. In this work, we will focus on probably the simplest generalization of a spherically symmetric black hole of Einstein-Maxwell theory, which includes the cosmological constant term, namely, the Reissner-Nordstr\"om-Anti de Sitter (RN-AdS) black hole.

An interesting consequence of the recent observations is the possibility of measuring the parameters that determine the structure of the black hole shadow, namely, the horizon, photon sphere, and horizon radii. 
In particular, in recent years, it has been concluded that observables such as the shadow radius are optimal for describing the thermodynamic properties of black holes, as they accurately reflect phenomena like phase transitions. This insight has led to the development of shadow thermodynamics, a formalism that analyzes the thermodynamic properties of black holes based on their shadows. This approach has already been explored in several alternative theories of gravity \cite{ZhangGuo, Luo, GuoLi, Belhaj, WangRuppeiner, Li, Kumar, Du, Zheng, He, ref20, ref21,adler2022cosmological,claudel2001geometry}.

On the other hand, differential geometric methods have been applied to study thermodynamic systems from a different perspective. In particular, 
the approach of thermodynamic geometry uses Hessian metrics to represent the equilibrium space of thermodynamic systems as Riemannian manifolds \cite{amari2012differential,weinhold1976metric,ruppeiner1979thermodynamics}. For instance, Ruppeiner metric has been used intensively to study black hole thermodynamics. Recently, the formalism of geometrothermodynamics (GTD) \cite{quevedo2007geometrothermodynamics} has been proposed to incorporate the Legendre invariance of classical thermodynamics into the geometric description of the equilibrium space. In fact, numerous relevant and recent studies utilize the thermodynamic geometry approach to analyze phase transitions, criticality, and the microstructure of a wide variety of black holes from different theories \cite{ref22,ref23,ref24,ref25,ref26,ref27,ref28,ref29,ref30,ref31,ref32,ref33,ref34,ref35,ref36,ref37,ref38,ref39,ref40,ref41,ref42,ref43,ref44,ref45,ref46,ref47,ref48,ref49,ref50,ref51,ref52,ref53,ref54,ref55,ref56,ref57,ref58,ref59}
. In this work, we will follow the approach of GTD to study the properties of the RN-AdS black hole. 

An important consequence of the use of thermodynamic geometry in black hole physics is that the thermodynamic curvature seems to contain indications about the microstructure of the system. 
Specifically, in \cite{ref14}, it is shown that the Ruppeiner scalar curvature of the Schwarzschild-AdS black hole  is negative, indicating attractive interaction between its constituents.  Moreover, the phase transitions and microstructure of RN-AdS black holes have been examined using thermodynamic geometry from different perspectives in \cite{ref15, ref16, ref17, ref18, ref19, WangRuppeiner, ref20, ref21}. In particular, in \cite{WangRuppeiner, ref20, ref21}, it was shown that using Ruppeiner geometry, the black hole shadow helps analyze the phase structure and microstructure of RN-AdS black holes.
Also, in \cite{ref13}, the thermodynamic geodesics of 4D asymptotically AdS black holes were studied using the GTD type II metric. This study concludes that the turning behavior or incompleteness of geodesics in GTD type II geometry can indicate phase transitions in these black holes.

The main goal of the present work is to study the main thermodynamic properties of the RN-AdS black hole, including shadow thermodynamics, phase transitions, and microstructure, using the formalism of GTD. This paper is organized as follows. In Sec. \ref{sec:shadow}, we review the main aspects of the RN-AdS black hole shadow, emphasizing the role of the observer location.
In Sec. \ref{sec:shtd}, we apply the ideas of shadow thermodynamics to the particular case of the RN-AdS black hole. In particular, we investigate the behavior of the main thermodynamic quantities in terms of the shadow radius, which constitutes the essential link between phase transitions and shadows.  
In Sec. \ref{sec:gtd}, we apply the formalism of two- and three-dimensional GTD to study in detail the phase transition structure of the RN-AdS spacetime, using the hypothesis of quasi-homogeneous thermodynamics and, consequently, considering the curvature radius as a thermodynamic variable. 
Furthermore, in Sec. \ref{sec:shgtd}, we analyze the GTD curvature from the point of view of the shadow parameters
and establish the possibility of inferring thermodynamic properties from the observation of shadows. Finally, in Sec. \ref{sec:mic}, we summarize our results and comment on possible future works.

\section{Shadow of the RN-AdS black hole}
\label{sec:shadow}

The RNAdS spacetime is a static, spherically symmetric, electrovacuum solution of Einstein-Maxwell field equations with cosmological constant. 
The corresponding line element can be represented as 
\begin{equation}
d s^2 = f(r) d t^2 + f(r)^{-1} d r^2 + r^2 (d\theta^2 + \sin^2\theta d\phi^2) , \label{elementolinea1}
\end{equation}
with 
\be 
f(r) = 1 - \frac{2M}{r} + \frac{Q^2}{r^2} + \frac{r^2}{l^2},
\label{lapse}
\ee 
where $M$ is the mass parameter, $Q$ the electric charge, and $l$ the curvature radius related to the cosmological constant by $\Lambda = -3 /l^2$.

\begin{figure}
    \centering
    \includegraphics[scale=0.08]{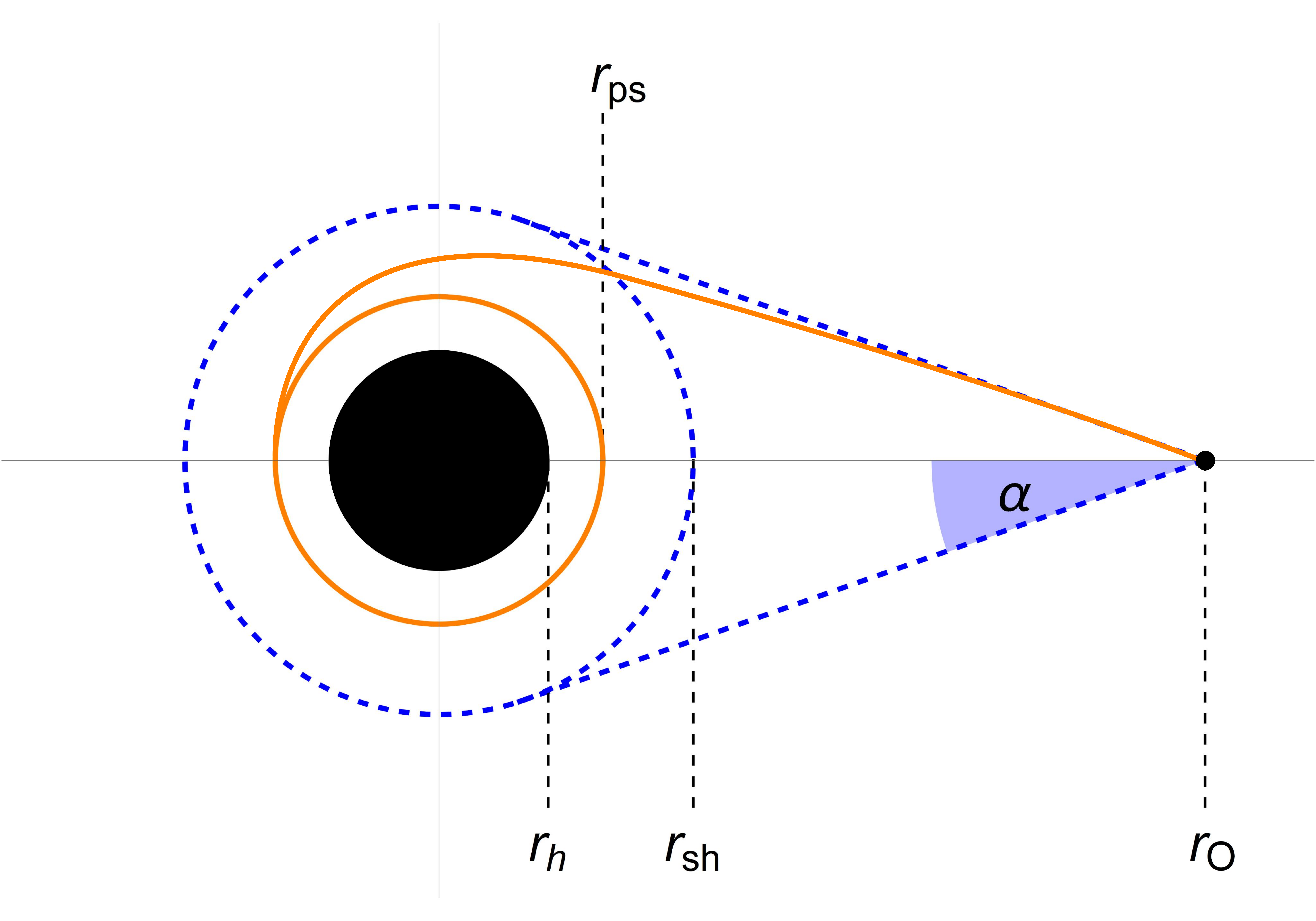}
    \caption{Schematic representation of the black hole shadow. The observer location $r_O$, the escape angle $\alpha$ and the radii of the horizon $r_h$, photon sphere $r_{ps}$, and shadow $r_{sh}$ are shown explicitly.   
    }
    \label{fig1}
\end{figure}

A shadow is a dark contour observed in the sky, resulting from the absorption of light rays by a black hole.  
In the case of spherically symmetric spacetimes, the shadow is essentially characterized by the shadow radius $r_{sh}$. Furthermore, the boundary of the shadow is determined by the photon sphere of radius $r_{ps}$, which is located outside the event horizon with radius $r_h$ and composed of photons moving along unstable circular orbits. Due to the light deflection caused by gravity,  the shadow radius is always larger than the radius of the photon sphere (see Fig. \ref{fig1}).

In the case of a spacetime described by the above line element, the  radius of the photon  sphere, $r_{ps}$, is determined by the solution to the following implicit equation \cite{Vagnozzi,Perlick}
\begin{equation}
f\left(r_{ps}\right)-\frac{1}{2} r_{ps} f^{\prime}\left(r_{ps}\right)=0 ,
\label{photonsphere}
\end{equation}
where the prime denotes differentiation with respect to the spatial coordinate $r$. 
In the case of the RN-AdS black hole presented above, the radius of the photon sphere  is given by
\begin{equation}
r_{ps} = \frac{1}{2}\left(3 M+\sqrt{9 M^2-8 Q^2}\right),
\end{equation}
which is independent of the parameter $l$.

Additionally, the shadow radius can expressed in terms of the escape radius and the location of the observer, see Fig. \ref{fig1}. Then, for large values of the observer radius, it can be shown that \cite{Perlick}
\begin{equation}
r_{sh}=r_O \tan \alpha \approx r_O \sin \alpha=r_{ps} \sqrt{\frac{f\left(r_O\right)}{f(r_{ps})}}\ .
\label{shadow}
\end{equation}
Then, for the RN-AdS black hole we obtain
\begin{equation}
r_{sh}=\frac{r_{ps}^2}{r_O} \sqrt{\frac{r_O^4+l^2\left[Q^2+r_O(r_O-2 M)\right]}{r_{ps}^4+l^2\left(Q^2+r_{ps}(r_{ps} -2 M)\right)}}.
\end{equation}
Next, by employing the mass equation 
\begin{equation}
M=\frac{Q^2}{2 r_{h}}+\frac{r_{h}}{2}+\frac{r_{h}^3}{2 l^2}
\end{equation}
in the above result, we can express the shadow radius in terms of the horizon radius as
\begin{equation}
r_{sh}(r_{h})=\frac{r_{psh}^2}{r_O} \sqrt{\frac{(r_{h}-r_O)\left[l^2\left(r_{h} r_O-Q^2\right)+r_{h} r_O\left(r_{h}^2+r_{h} r_O+r_O^2\right)\right]}{(r_{h}-r_{psh})\left[r_{h} r_{psh}\left(l^2+r_{h}^2+r_{psh}(r_{h}+r_{psh})\right)-l^2 Q^2\right]}} \ ,
\label{shadowofrh}
\end{equation}
with
\begin{equation}
r_{psh}\equiv r_{ps}(r_{h})=\frac{1}{2}\left(\frac{3 r_{h}^3}{2l^2}+\frac{3\left(Q^2+r_{h}^2\right)}{2 r_{h}}+\sqrt{\frac{9\left[r_{h}^4+l^2\left(Q^2+r_{h}^2\right)\right]^2}{4l^4 r_{h}^2}-8 Q^2}\right)
\end{equation}
In the limit $r_{h} \rightarrow r_O$, we confirm that $r_{sh} =0$ and in the limit  $r_{ps} \rightarrow r_O$, it follows that  $r_{sh} = r_O$.\\

In the sections below, we will  need the relation $r_{h}(r_{sh})$, which, however, cannot be obtained analytically. Therefore, we will apply a numerical analysis. The shadow boundary observed at $r_O$ can be obtained by applying a stereographic projection in terms of the celestial coordinates $(x, y)$ 
\cite{Perlick, ZhangGuo, Luo, GuoLi, Belhaj, WangRuppeiner, Li, Kumar, Du, Zheng, He}
\begin{align}
& x=\lim _{r \rightarrow \infty}-\left.r^2 \sin \theta_0 \frac{d \phi}{d r}\right|_{\theta_0=\frac{\pi}{2}}, \label{Projection1} \\
& y=\lim _{r \rightarrow \infty} r^2 \frac{d \theta}{d r} . \label{Projection2}
\end{align}
This projection determines the shape of the shadow and, as we will see later, can be used to analyze the behavior of thermodynamic properties on the shadow's profile \cite{WangRuppeiner}.


\section{Shadow thermodynamics of the RN-AdS black hole}
\label{sec:shtd}

The main ingredient  of black hole
thermodynamics is the Bekenstein-Hawking formula \cite{bekenstein1973black}, which relates the  entropy $S$ with the horizon area, $S=\frac{1}{4} A_h$. 
In the case of spherically symmetric spacetimes, the entropy becomes $S=\pi r_h^2$. Furthermore, this definition is consistent with the expressions for the Hawking temperature \cite{hawking1974black}
\begin{equation}
\label{temp_charged}
T=\left.\frac{f^{\prime}(r)}{4 \pi}\right|_{r=r_h}=\frac{3 r_h^4+l^2\left( r_h^2-Q^2\right)}{4l^2 \pi r_h^3}.
\end{equation}
From here,  we can determine the heat capacity at constant $l$ and $Q$, according to the definition 
\begin{equation}
C_{lQ}=T\left(\frac{\partial S}{\partial T}\right)_{l,Q}=\frac{2 \pi r_h^2\left(l^2 r_h^2+3 r_h^4-l^2 Q^2\right)}{3 l^2 Q^2-l^2 r_h^2+3 r_h^4}.
\end{equation}
The temperature should be a positive quantity, and its minimum value $T=0$ is reached at 
\begin{equation}
r_{h}(T=0) =\frac{1}{\sqrt{6}} \sqrt{\sqrt{l^2\left(l^2+12 Q^2\right)}-l^2}.
\label{extremeT}
\end{equation}
Interestingly, the heat capacity diverges exactly at those locations where the temperature is extremal, i.e., for  $\partial T/\partial r_h =0 $, which correspond to the locations 
\begin{equation}
r_{h}(T=T_{extr}) =\frac{1}{\sqrt{6}} \sqrt{l^2 \pm \sqrt{l^2\left(l^2-36 Q^2\right)}},
\label{criticalT}
\end{equation}
so that 
\begin{equation}
T_{min,max}=\frac{\sqrt{6}\left(l^2-12 Q^2 \pm l \sqrt{l^2-36 Q^2}\right)}{2 \pi\left(l^2 \pm l \sqrt{l^2-36 Q^2}\right)^{3 / 2}}.
\label{Tminmax}
\end{equation}
This expression allows us to identify the  critical value $l_{c}=6Q$ from which we can determine the remaining critical parameters \cite{Mann, Belhaj, WangRuppeiner}
\begin{equation}
r_{hc}=\sqrt{6}Q, \quad M_{c}=\frac{2}{3}\sqrt{6}Q, \quad T_{c}=\frac{1}{3\sqrt{6} \pi Q}, \quad r_{psc}=(2+\sqrt{6})Q,
\end{equation}
and
\begin{equation}
r_{shc}=\frac{1}{2 r_O} \sqrt{\frac{1}{23}(29+12 \sqrt{6})\left(12 Q^2\left(3 Q^2-4 \sqrt{6} Q r_O+3 r_O^2\right)+r_O^4\right)}.
\end{equation}

\begin{figure}
\includegraphics[width=0.5\linewidth]{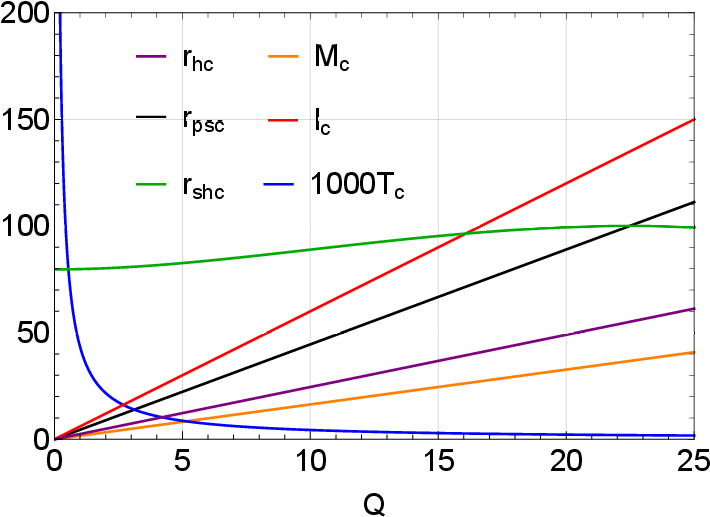}
 \centering
    \caption{Behavior of the critical quantities as functions of the black hole electric charge $Q$ (with $r_O=100$). \label{Figure2}}
\end{figure}

In Fig. \ref{Figure2}, we illustrate the behavior of the above critical quantities as  functions of the black hole electric charge $Q$. We can see that all these critical quantities, except for $T_c$,  grow as $Q$ increases. 
In particular, the critical value $r_{shc}$ is not linearly proportional to $Q$. However, taking as a lower bound $Q \rightarrow 0$ and as an upper bound the value $r_{shc} \rightarrow r_O$, we can infer some constraints for $r_{shc}$ and $Q$, namely,
\begin{equation}
 \frac{1}{2} \sqrt{\frac{1}{23} \left(29 + 12 \sqrt{6}\right)}  r_O  < r_{shc}<r_O,\quad \quad\text{for}\quad \quad\ 0<Q<\frac{1}{2}\left(\sqrt{6}-2\right)r_O .
 \label{restriccion2}
\end{equation}
This is approximately
\begin{equation}
\hspace{1.2cm}0.7967 r_O < r_{shc}<r_O, \quad\quad\text{for} \quad\quad 0<Q<0.2247r_O
 \label{restriccion3}
\end{equation}
The critical value of the shadow radius, $r_{shc}$, can be interpreted as the value at which the shadow of the black hole represents a 
second-order phase transition, reaching the bounds given in 
Eq.\eqref{restriccion2}. Consequently, we have identified a constraint on the possible values of the critical shadow radius, $r_{shc}$, and the electric charge, $Q$, for an RN-AdS black hole to undergo a second-order phase transition, which depends on the distance to the observer $r_O$.  This is in accordance with Eqs.\eqref{restriccion2} and \eqref{restriccion3}, which provide these constraints. This means that an RN-AdS black hole with a shadow radius $r_{sh} > 0.7967 r_O$ could undergo a second-order phase transition before reaching $r_{sh}=r_O$. 

In general, for Schwarzschild-AdS black holes, the values of the minima of $T$ and the divergences of $C_l$, are reached for  
\begin{equation}
r_h=l/\sqrt{3},\quad \quad r_{ps}=2l/\sqrt{3}, \quad  \quad r_{sh}=\frac{2 r_O}{3^{3 / 4} \sqrt{5}} \sqrt{3 \sqrt{3}+\frac{l^2(3 \sqrt{3} r_O-4l)}{r_O^3}}.
\label{criticSchwarzschild}
\end{equation}

\begin{figure}
\begin{minipage}[t]{0.47\linewidth}
 \centering
\hspace{1cm}
\includegraphics[width=1\linewidth]{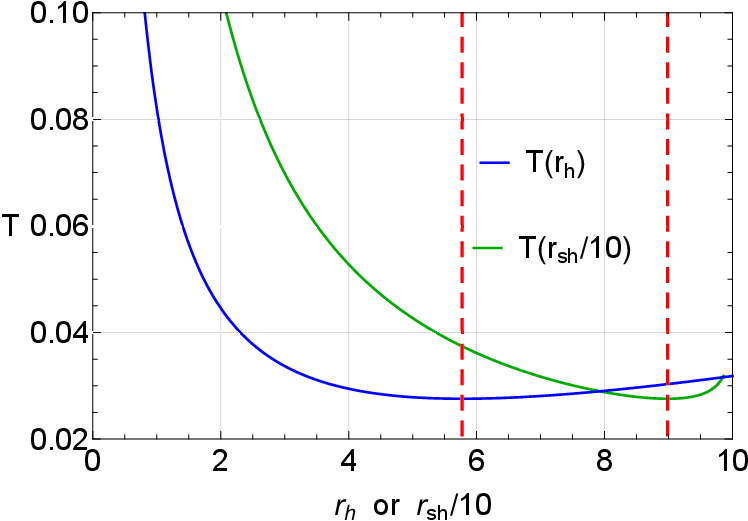}
 (a)\hspace{5cm}
\end{minipage}%
\hfill%
\begin{minipage}[t]{0.49\linewidth}
 \centering
\hspace{1cm}
\includegraphics[width=1\linewidth]{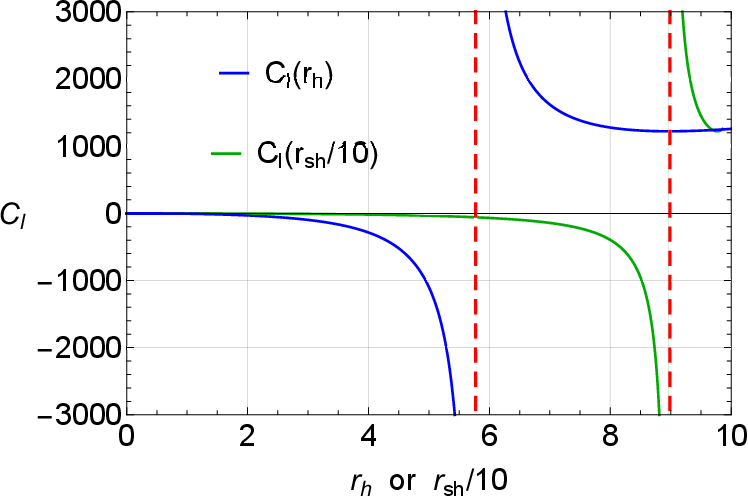}
(b)\hspace{5cm}
\end{minipage}
    \caption{(a) Hawking temperature $T$ and (b) heat capacity $C_l$ as  functions of the horizon radius $r_{h}$ and the scaled shadow radius $r_{sh}/10$, for a Schwarzschild-AdS black hole (with $r_O=100$ and $l=10$). The red dotted lines indicate the values of the minima of $T$ and the divergences of $C_l$.
 \label{Figure1}}
\end{figure}

In Fig. \ref{Figure1}, we present $T$ and $C_l$ as functions of the horizon radius $r_{h}$ and the scaled shadow radius $r_{sh}/10$ for a Schwarzschild-AdS black hole. The values of the minima of $T$ and the divergences of $C_l$ are highlighted with red dotted lines. The minima of $T$ matches the divergences of $C_l$.  This system reveals that both parameters, $r_{h}$ and $r_{sh}$, indicate a second-order phase transition in $C_l$. Using 
Eqs.\eqref{criticSchwarzschild}, the heat capacity shows a divergence at $r_{h} \approx 5.7735$ and $r_{sh} \approx 89.8546$ (with $r_O = 100$ and $l = 10$) and exhibits both positive and negative phases. Therefore, the black hole is stable only in the region where the positive phase exists for small values of these critical values of $r_{h}$ and $r_{sh}$.

\begin{figure}
\begin{minipage}[t]{0.47\linewidth}
 \centering
\hspace{1cm}
\includegraphics[width=1\linewidth]{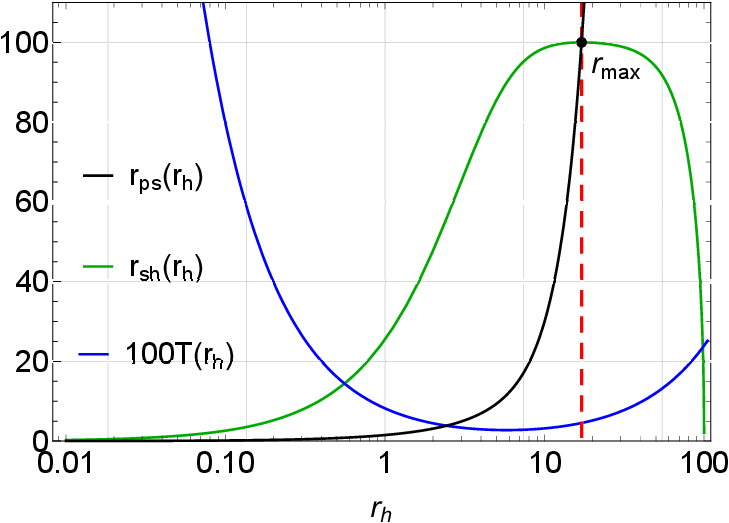}
 (a)\hspace{5cm}
\end{minipage}%
\hfill%
\begin{minipage}[t]{0.49\linewidth}
 \centering
\hspace{1cm}
\includegraphics[width=1\linewidth]{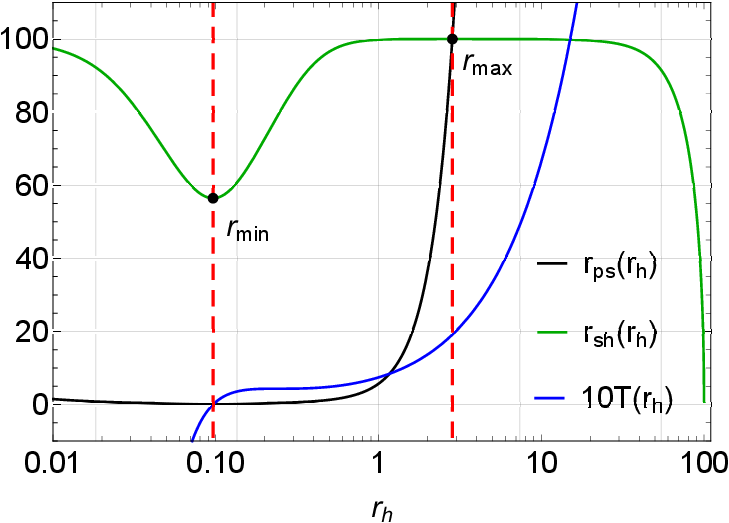}
(b)\hspace{5cm}
\end{minipage}
    \caption{(a) Scaled Hawking temperature $100T$, photon sphere radius $r_{ps}$, and shadow radius $r_{sh}$ as  functions of the  horizon radius $r_{h}$ for the Schwarzschild-AdS black hole (with $r_O=100$ and $l=10$). (b) Scaled Hawking temperature $10T$, photon sphere radius $r_{ps}$, and shadow radius $r_{sh}$ as functions of the  horizon radius $r_{h}$ for the RN-AdS black hole (with $r_O=100$, $Q=0.1$ and $l=l_{c}$). The red dotted lines indicate the possible limit values for $r_{sh}$.
  \label{Figure3}}
\end{figure}

As shown in Fig. \ref{Figure3}, the three solid curves represent the scaled Hawking temperature, the photon sphere radius, and the shadow radius as functions of the horizon radius $r_{h}$. The local extremum points are indicated by the red dotted lines. For the Schwarzschild-AdS black hole, there is one local extremum  marked as $r_{max}$. Only the segment where $r_{sh} < r_{max}$ is relevant to reflect the phase transition of the black hole, as the remaining segment is non-physical. This is because the shadow radius for $r_{sh} > r_{max}$ is no longer applicable, since the observer is always assumed to be located outside the photon sphere. In the case of the RN-AdS black hole, there are two local extremum points, $r_{min}$ and $r_{max}$, dividing the shadow curve into three parts. Only the segment $r_{min} < r_{sh} < r_{max}$ is relevant for reflecting the phase transition of an RN-AdS black hole. The other two segments are non-physical because $r_{sh} < r_{min}$ corresponds to $T<0$, and $r_{sh} > r_{max}$ is not applicable, as the observer is always considered to be situated beyond the photon sphere \cite{WangRuppeiner}.

Examining the local extremum points, we have $r_{max}=r_{O}=r_{ps}=r_{sh}$. To find $r_{h}$ at which $r_{max}$ is reached, we can start analytically from the local maximum of $r_{sh}$ or $r_{ps}$ upon reaching $r_O$, {this happens exactly in}
\begin{equation}
 r_h=\frac{1}{6} \sqrt{\mathcal{A}}+\frac{1}{2} \sqrt{\frac{4l^2\left(2 Q^2+r_O^2\right)}{r_O \sqrt{\mathcal{A}}}-\frac{4l^2}{3}-\frac{\mathcal{B}}{1458^{1 / 3} r_O}-\frac{2^{1 / 3}\left(l^4+12l^2 Q^2\right) r_O}{\mathcal{B}}},
\label{criticrhofrO}
\end{equation}
where
\begin{align}
& \mathcal{A}=-6l^2+\frac{\mathcal{B}}{2^{1 / 3} r_O}+\frac{1458^{1 / 3} l^2\left(l^2+12 Q^2\right) r_O}{\mathcal{B}},  \\
& \mathcal{B} =3\Big\{ 2l^4 r_O\left[24 Q^4+\left(l^2-12 Q^2\right) r_O^2+6 r_O^4\right]  \notag \\
 & \hspace{0.7cm}  + 4 \sqrt{3} \sqrt{l^6 r_O^2\left[48 Q^6-l^2 r_O^2\left(r_O^2-4 Q^2\right)\right]\left(l^2 Q^2-l^2 r_O^2-3 r_O^4\right)} \Big\}^{1 / 3}. 
\end{align}

In Fig. \ref{Figure3}, we show 
the scenarios for a Schwarzschild-AdS black hole with $r_{max}=r_{O}=r_{ps}=r_{sh}=100$, $l=10$, reaching $r_{max}$ at  $r_{h}=17.0553$, and for an RN-AdS black hole with  $r_{max}=r_{O}=r_{ps}=r_{sh}=100$, which is reached  when $r_{h}=2.8428$, employing Eq.\eqref{criticrhofrO}.
Using Eq.\eqref{extremeT} in \eqref{shadowofrh}, we find that $r_{sh}=r_{min}=56.4483$ (with $Q=0.1$ and $l=l_{c}$). Thus, we have established a lower bound for $r_{sh}$. For instance, for $r_O=100$, $Q=0.1$ and $l=l_{c}$, the thermodynamic shadow analysis for the RN-AdS black hole is valid if 
\begin{equation}
0.2076 < r_{ps} < 100 \quad\quad\text{and} \quad\quad 56.4483 < r_{sh} < 100.
\end{equation}
Note that the constraints found for the critical shadow radius, $r_{shc}$, and the electric charge, $Q$, from Eq.\eqref{restriccion2} are compatible with the restriction of the previous expressions.

\begin{figure}
\begin{minipage}[t]{0.47\linewidth}
 \centering
\hspace{1cm}
\includegraphics[width=1\linewidth]{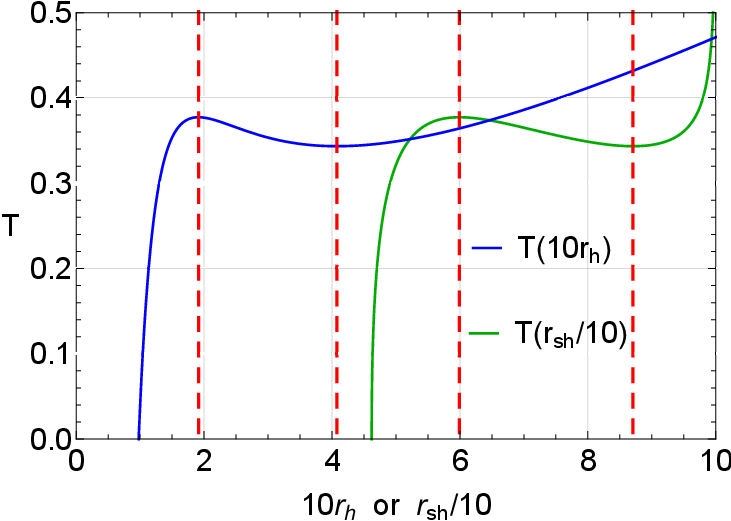}
 (a)\hspace{5cm}
\end{minipage}%
\hfill%
\begin{minipage}[t]{0.49\linewidth}
 \centering
\hspace{1cm}
\includegraphics[width=1\linewidth]{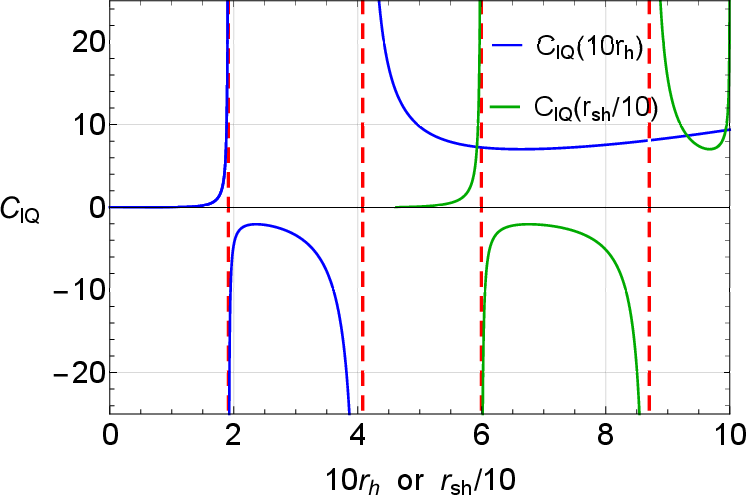}
(b)\hspace{5cm}
\end{minipage}
    \caption{(a) Hawking temperature $T$ and (b) Heat capacity $C_{lQ}$ as a function of the scaled horizon radius $10r_{h}$ and the scaled shadow radius $r_{sh}/10$ (with $r_O=100$, $Q=0.1$ and $l=1.3l_{c}$). The red dotted lines indicate the values of the minima and maxima of $T$ and the divergences of $C_{lQ}$. \label{Figure4}}
\end{figure}

\begin{figure}[H]
\begin{minipage}[t]{0.47\linewidth}
 \centering
\hspace{1cm}
\includegraphics[width=1\linewidth]{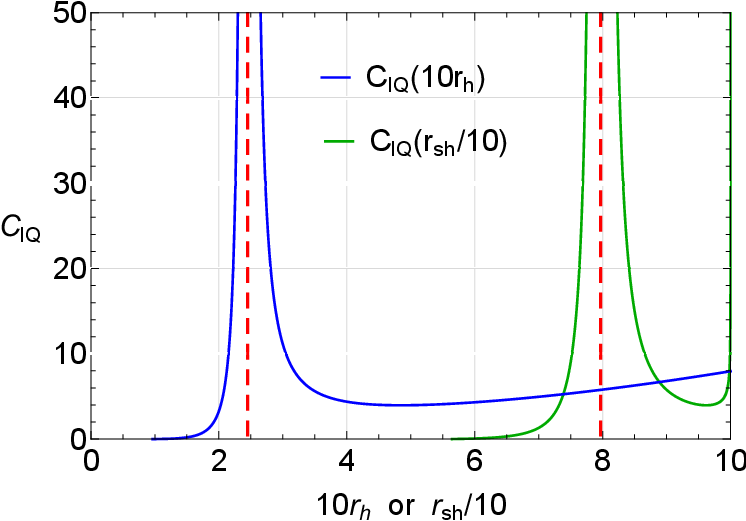}
 (a)\hspace{5cm}
\end{minipage}%
\hfill%
\begin{minipage}[t]{0.49\linewidth}
 \centering
\hspace{1cm}
\includegraphics[width=1\linewidth]{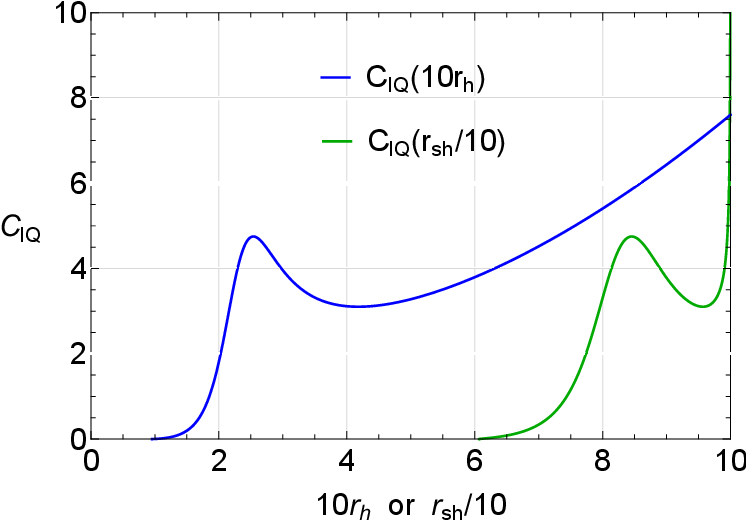}
(b)\hspace{5cm}
\end{minipage}
    \caption{Heat capacity $C_{lQ}$ as a function of the scaled horizon radius $10r_{h}$ and the scaled shadow radius $r_{sh}/10$ (with $r_O=100$ and $Q=0.1$). (a) for $l=l_{c}$ and (b) for $l=0.9l_{c}$. The red dotted lines indicate the divergences of $C_{lQ}$. \label{Figure5}}
\end{figure}

In Figs. \ref{Figure4} and \ref{Figure5}, we present $T$ and $C_{lQ}$ as functions of the scaled horizon radius $10r_{h}$ and the scaled shadow radius $r_{sh}/10$. The minimum and maximum values of $T$, along with the divergences of $C_{lQ}$, are determined using 
Eq.\eqref{criticalT} for $r_h$. Substituting \eqref{criticalT} into  \eqref{shadowofrh} yields the corresponding values for $r_{sh}$. For $r_O=100$, $Q=0.1$, and $l=1.3l_{c}$, the case of the Fig.\ref{Figure4} when $l>l_{c}$, these quantities are attained at $r_{h}=0.1913$ and $r_{h}=0.4077$ for the horizon radius, and $r_{sh}=59.9239$ and $r_{sh}=87.0504$ for the shadow radius. In the cases of 
Fig.  \ref{Figure5}, the heat capacity $C_{lQ}$ is shown on the left for $l=l_{c}$ and on the right for $l<l_{c}$. When $l=l_{c}$ the heat capacity has a divergence at $r_{h}=0.2449$ and $r_{sh}=79.6705$. And when $l<l_{c}$, the behavior of $C_{lQ}$ does not show any divergence.

In the following sections, we will delve deeper into these results. By applying the theory of GTD to shadows, we will analyze the shadow of the RN-AdS black hole to reveal and discuss its critical thermodynamic behavior.

\section{Geometrothermodynamics }
\label{sec:gtd}

One of the main ingredients of GTD is Legendre invariance, which in ordinary equilibrium thermodynamics means that the properties of a system do not depend on the choice of thermodynamic potential used for its description. GTD incorporates Legendre invariance into the formalism by introducing the auxiliary phase space ${\cal T}$, which is a $2n+1$ dimensional manifold with metric $G_{AB}$, $A,B=0,1,...,2n$, where $n$ is the number of thermodynamic degrees of freedom. For concreteness, we introduce the set of coordinates $Z^A = \{ \Phi, E^a, I_a\}$ with $a=1,...,n$. Then, the line elements $G =G_{AB}dZ^AdZ^B$  of the Legendre invariant metrics of ${\cal T}$ can be expressed as 
\begin{equation}
    G^{I/II}=\left(d\Phi-I_aE^a\right)^2+(\xi_{ab}E^aI^b)(\chi_{cd}dE^cdI^d),
\end{equation}
\begin{equation}
    G^{III}=\left(d\Phi-I_aE^a\right)^2+\sum_{a=1}^{n}\xi_{a}(E_aI_a)^{2k+1}dE^adI^a,
\end{equation}
where $\delta^c_a=diag(1,1, \ldots , 1)$, $\eta^c_a=diag(-1, \ldots , 1)$, and $\xi_{a}$ are real constants. Furthermore,  $\chi_{cd} = \delta_{ac}$ for $G^I$ and $\chi_{cd} = \eta_{cd}$ for $G^{II}$, $\xi_{ab}$ is a diagonal ($n\times n$) real matrix, and $k$ is an integer. 

The equilibrium space ${\cal E}$ with coordinates $E^a$ is an $n-$dimensional subspace of ${\cal T}$ defined by the mapping $\varphi: {\cal E} \rightarrow {\cal T}$ so that the coordinates $Z^A$ become functions of $E^a$, i.e., 
$Z^A \rightarrow Z^A(E^a) =\{ \Phi (E^a), E^a, I_a(E^a)\}$, where $\Phi=\Phi(E^a)$ is the fundamental equation of the thermodynamic system \cite{callen1998thermodynamics}. Furthermore,  ${\cal E}$ can be endowed with a Riemannian metric $g_{ab}$, which is determined by  the pullback $\varphi^*(G) = g= g_{ab}dE^adE^b$. Then, the 
corresponding induced metrics on the equilibrium space are:
\begin{equation}
    g^I_{ab}=\beta_\Phi \Phi \delta^c_a\frac{\partial^2 \Phi}{\partial E^b \partial E^c}, \label{g1}
\end{equation}
\begin{equation}
    g^{II}_{ab}=\beta_\Phi \Phi \eta^c_a\frac{\partial^2 \Phi}{\partial E^b \partial E^c}, \label{g2}
\end{equation}
\begin{equation}
    g^{III}=\sum_{a=1}^{n}\beta_a\left(\delta_{ad}E^d\frac{\partial \Phi}{\partial E^a}\right)^{2k+1}\delta^{ab}\frac{\partial^2 \Phi}{\partial E^b \partial E^c} dE^adE^c.\label{g3}
\end{equation}
To obtain the components of the metrics $g^I$ and $g^{II}$, we have chosen $\xi_{ab}=\beta_{ab}=diag(\beta_1,...., \beta_n)$, where $\beta_a$ are the quasi-homogeneous coefficients determined by the condition $\Phi(\lambda^{\beta_a} E^a = \lambda^{\beta_\Phi} \Phi(E^a)$ with $\lambda$ being a positive real constant. Moreover, we have used quasi-homogeneous Euler identity, $\sum_a \beta_a E^a I_a = \sum_a \beta_a E^ a \frac{\partial \Phi}{\partial E^a} = \beta_\Phi \Phi$ \cite{quevedo2019quasi}. The explicit expressions for the above metrics can be further analyzed by fixing the number of thermodynamic degrees of freedom $n$. In appendices \ref{app:2dof} and \ref{km}, we study in detail the cases $n=2$ and $n=3$. In the forthcoming subsections, we analyze both the Schwarzschild-AdS and RN-AdS black hole configurations in the context of GTD.

\subsection{Schwarzschild-AdS black hole }

We now consider the Schwarzschild-AdS black hole solution as a quasi-homogeneous thermodynamic system. The fundamental equation follows from the condition that the lapse function (\ref{lapse}) vanishes at the horizon. Thus, we obtain 
\begin{equation}
     M(S,l)=\frac{1}{2}\frac{\sqrt{S}}{\sqrt{\pi}}\Big[1+\frac{S}{\pi l^2}\Big].
    \label{fundame1}
\end{equation}
The Hawking temperature is given by Eq.(\ref{temp_charged}), when the charge is set to zero, i.e., \begin{equation}
\label{tempe_schwar}
T=\frac{3 r_h^2+l^2}{4\pi l^2r_h}.
\end{equation}
Performing the rescaling of the extensive variables, it is easy to see that Eq.(\ref{fundame1}) is a quasi-homogeneous function of degree $\beta_M$, 
if the condition
\begin{equation}
    \beta_l=\frac{1}{2}\beta_S,\quad \beta_M\equiv \beta_l, \label{escal}
\end{equation}
is imposed.  With this condition, it is trivial to check that the Euler identity is fulfilled, which can be used to simplify the calculations as shown in Appendix \ref{app:2dof}.  Then, from Eqs.(\ref{2g11})-(\ref{g233}), we obtain the line elements 
\be
g^I = \frac{\beta_M M}{\pi^{3/2}l^2}\left(\frac{3S-\pi l^2}{8 S^{3/2}} dS ^2
- \frac{3S^{1/2}}{l}dM dl + \frac{3S^{3/2}}{l^2} dl^2\right) ,
\ee
\be
g^{II} = \frac{\beta_M M}{\pi^{3/2}l^2}\left(-\frac{3S-\pi l^2}{8 S^{3/2}} dS ^2
+ \frac{3S^{3/2}}{l^2} dl^2\right) ,
\ee
\be
g^{III} = \frac{\beta_M }{\pi^{3/2}l^3}\left( 
\frac{9S^2 -\pi^2 l^4}{16\pi^{3/2} S l } dS^2 - \frac{3}{2} MS^{1/2} dS dl 
-\frac{3S^3}{\pi^{3/2}l^3} dl^2\right) ,
\ee
where we have used the Euler identity, $\beta_S S M_{,S} + \beta_l l M_{,l} = \beta_M M $, the relationships between the quasi-homogeneity coefficients 
(\ref{escal}), and the condition $k=0$ to simplify the expression for $g^{III}$. 
As explained in Appendix \ref{app:2dof}, to guarantee that the above three metrics represent the same thermodynamic system, it is necessary to consider the curvature singularities of all the metrics  simultaneously. Then, according to Eqs.(\ref{c1})--(\ref{c3}), the singularities are determined by the conditions 
  \begin{align}
      I&:M_{,SS} M_{,ll}- (M _{,Sl})^2= -\frac{3(3S+\pi l^2)}{8 \pi^3 l^6}=0, \label{c1schw}\\
       II&:M_{,SS} M _{,ll}= 
\frac{3(3S - \pi l^2)}{8 \pi^3 l^6} = 
       0,\\
        III&: M _{,Sl}= -\frac{3S^{1/2}}{2\pi^{3/2}l^3} =  0.\label{c3schw}\
\end{align}
Condition $I$ and $III$ cannot be satisfied in general, whereas the singularity $II$, located at $S=\pi l^2 / 3$, implies that for a particular value of the curvature radius, there always exists a positive value of the entropy for which a curvature singularity exists. 

\begin{figure}[H]
    \centering
    \includegraphics[width=0.6\linewidth]{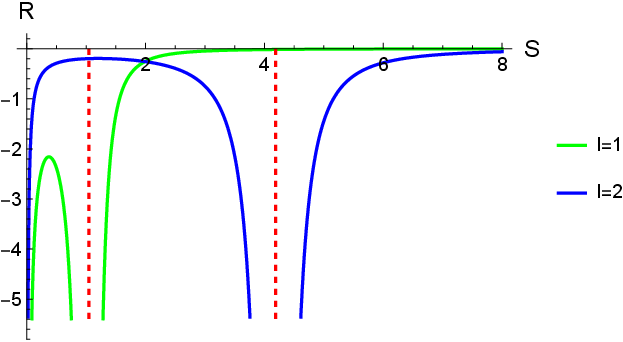}
    \caption{Behavior of the Ricci scalar $R^{II}$ of the Schwarzschild-AdS black hole in terms of the entropy $S$ for different values of the radius of curvature $l$ and $\beta_M=1$.}
        \label{figRII}
\end{figure}

\begin{figure}[H]
\begin{minipage}[t]{1\linewidth}
 \centering
\hspace{1cm}
\includegraphics[width=0.8\linewidth]{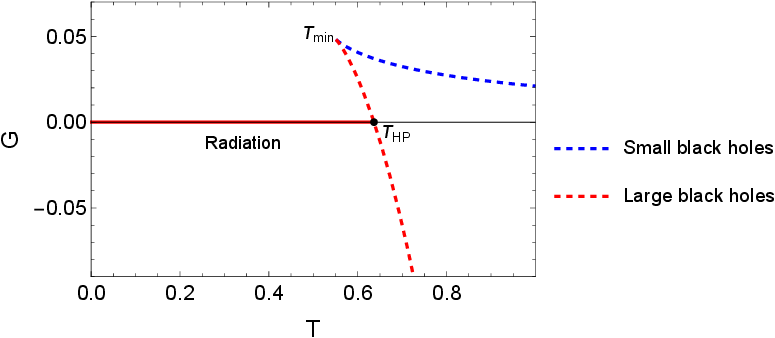}
(b)\hspace{5cm}
\end{minipage}
    \caption{Gibbs Free energy of the Schwarzschild-AdS black hole is displayed  for fixed $l=0.5$. At $T_{min}\approx 0.55$, we have a discontinuity in the first derivative of $G$ indicating a possible phase transition between small/large black holes. For  $T>T_{HP}$ the lower branch of large black holes has negative  free energy and corresponds to the globally thermodynamically preferred state. At $T=T_{HP}$, we observe a discontinuity in the first derivative of the radiation/black hole free energy, characteristic of first order phase transitions.}\label{freeenergies}
\end{figure}

To obtain a direct relation with the response functions of the thermodynamic system, we might use the general results obtained in Appendix \ref{km}. Thus, in the two-dimensional case the singularities conditions \cite{quevedo2023unified}, read
\begin{align}
    I&: \frac{T}{C_L\kappa_S}=0\label{gf1},\\
     II&: \frac{T\kappa_S}{C_l}=0,\\
      III&:\frac{1}{\alpha_{S}} =0,\label{gf3}
    \end{align}
where the heat capacities, and compressibility parameters are defined as
\begin{align}
C_{l}&=T\Big(\frac{\partial S}{\partial T}\Big)_{l}=\frac{2S(3S+\pi l^2)}{3S-\pi l^2},\label{heatcl}\\
C_{L}&=T\Big(\frac{\partial S}{\partial T}\Big)_{L}=-2S,\\
\kappa_S&=\Big(\frac{\partial l}{\partial L}\Big)_{S}=\frac{\pi^{3/2}l^4}{3S^{3/2}},\\
\alpha_S&=\Big(\frac{\partial l}{\partial T}\Big)_{S}=-\frac{2\pi^{3/2}l^3}{3S^{1/2}}.\\
\end{align}
From the above expressions, and from Eqs.(\ref{gf1})-(\ref{gf3}) is clear that the  singularity condition $II$ 
coincides with the divergence of the heat capacity $C_l$, indicating the presence of a phase transition. Moreover, the heat capacity $C_L$, the compressibility $\kappa_S$, and the coefficient of thermal expansion $\alpha_S$ are regular for $S\neq 0$. This traduces into singularities conditions $I$ and $III$ cannot be fulfilled in general. In fact, the singularity of $R^I$ coincides with the limit $T\rightarrow 0$, which is non-physical. The singularity associated with $R^{III}$, i.e., $M_{,Sl}=0$ is also non-physical because it implies the non-allowed thermodynamic limit $S\rightarrow 0$. It follows that the entire phase transition structure of the Schwarzschild-AdS black hole is determined by the behavior of the scalar $R^{II}$, which in this case can be expressed as 
\be
R^{II}=\frac{-\sqrt{\pi S}l^2\Big[13S^2+5\pi l^2 \Big]}{3\beta_M M^3\Big[3S-\pi l^2\Big]^2},  
\ee
and whose behavior is depicted in  Fig. \ref{figRII}.
To explore the phase transition predicted by $R^{II}$, we can compare it with the predictions of standard black hole thermodynamics theory, by analyzing the discontinuities in the derivatives of the Gibbs free energy. Following the standard approach in extended black hole thermodynamics, where the mass is regarded as the gravitational enthalpy \cite{kastor2009enthalpy}, in the canonical ensemble the Gibbs free energy is defined as:
\begin{align}
    G &= M - TS,
\end{align}
where $T=\partial M/\partial S$. Then, from Eq.(\ref{fundame1}), we obtain 
\begin{align}
G(T,l)=\frac{1}{4}r_h\Bigg[ 1-\frac{r_h^2}{l^2}\Bigg].\label{gibbs1}
\end{align}
Using Eq.(\ref{tempe_schwar}), we can solve analytically for $r_h(T,l)$, 
\begin{equation}
    r_h=\frac{l}{3}\Big[2\pi l T\pm \sqrt{4 \pi^2l^2T^2-3}\Big]. \label{rhgibbs}
\end{equation}
Notice that, for $T>T_{min}=\sqrt{3}/2\pi l$, we have two solutions, one that corresponds to small black holes ($r_h  <l/\sqrt{3}$), and one for large black holes ($r_h >l/\sqrt{3}$).  From Eq.(\ref{heatcl}) we can observe that $C_l$ is always negative for $r_h  <l/\sqrt{3}$ and positive for $r_h >l/\sqrt{3}$. This means that small Schwarzschild-AdS black holes cannot reach thermal equilibrium with its surroundings, implying that they are unstable configurations (see Fig. \ref{Figure1}). Conversely, large AdS black holes have positive heat capacity, and can be in stable equilibrium with the thermal radiation at fixed temperature \cite{hawking1983thermodynamics}. Inserting Eq.(\ref{rhgibbs}) in Eq.(\ref{gibbs1}), we obtain two branches for the free energy that meet at $T_{min}$ (see Fig. \ref{freeenergies}), exactly where the heat capacity $C_l$ diverges. Notably, $R^{II}$ diverges at $r_h=l/\sqrt{3}$, which corresponds to $T=T_{min}$. Therefore, GTD predicts correctly the phase transition for small/large Schwarzschild-AdS black holes. Furthermore, at temperatures higher than $T_{HP}=1/\pi l$, the configuration with a large black hole and thermal radiation has a lower free energy than the configuration with just thermal radiation and represents the globally preferred state. Therefore, at $T_{HP}=1/\pi l$, there is a first order phase transition between thermal radiation and large black holes, known as Hawking-Page transition \cite{hawking1983thermodynamics}. 

It is well-known that the stability properties of a black hole can depend on the statistical ensemble \cite{caldarelli2000thermodynamics, gibbons1977action}. However, in the thermodynamic limit, a change of ensemble can be simply performed as a Legendre transformation that acts on the thermodynamic
potential \cite{quevedo2014ensemble}. Consequently, the thermodynamic properties, like the phase transition structure of a black hole can depend on the choice of thermodynamic potential. Treating the black hole in the canonical ensemble has several issues. For example, a
small Schwarzschild AdS black hole ( $r_h  <l/\sqrt{3}$) has negative specific heat, energy fluctuations calculated in canonical ensemble
have formally negative variance \cite{gour1999schwarzschild}. Moreover, the canonical ensemble implies that the black hole is in thermal equilibrium with the
surrounding thermal bath \cite{hawking1983thermodynamics}. Thus the black hole mass remains constant because the density matrix of the canonical ensemble is constant in time \cite{gour1999schwarzschild}. However, the mass of a black hole formed by a gravitational collapse decreases in time because of the Hawking radiation \cite{hawking1975particle}, so a description of it via a canonical ensemble seems inadequate. Furthermore, in this work, we are treating the radius of curvature $l$ as thermodynamic variable. Hence, it would be more appropriate to consider an ensemble where $l$ is allowed to fluctuate.


\subsection{RN-AdS black hole}
The charged spherical AdS black hole solution with entropy $S=\pi r_h^2$, and temperature given by Eq.(\ref{temp_charged}), is described by the following fundamental equation
\begin{equation}
     M(S,Q,l)=\frac{1}{2}\Big[\frac{\sqrt{S}}{\sqrt{\pi}}+\frac{\sqrt{\pi}Q^2}{\sqrt{S}}+\frac{S^{3/2}}{\pi^{3/2} l^2}\Big].
    \label{fundame2}
\end{equation}
Performing the rescaling of the extensive variables,  $M(\lambda^{\beta_S} S,\lambda^{\beta_Q}Q,\lambda^{\beta_l}l)$, the fundamental equation is a quasi-homogeneous function of degree $\beta_M$, 
if the conditions
\begin{equation}
    \beta_l=\frac{1}{2}\beta_S,\quad \beta_Q\equiv \frac{1}{2}\beta_S,\quad \beta_M= \frac{1}{2}\beta_S, \label{escal2}
\end{equation}
are imposed. It is then trivial to check that the  Euler identity 
\be
\beta_ S S\frac{\partial M}{\partial S} +
\beta_ Q Q\frac{\partial M}{\partial Q} +
\beta_ l l\frac{\partial M}{\partial l} = \beta_M M ,
\ee
is fulfilled. Then, we can use the results presented in Appendix \ref{km}, which use explicitly the Euler identity.
According to Eqs.(\ref{g11})--(\ref{g33}), the line elements of the GTD metrics of the charged AdS black hole can be written as 
\be
 g^I = \frac{\beta_M M}{\pi^{3/2}}\left(\frac{3S^2-\pi S l^2+3\pi^2 Q^2l^2}{8 S^{5/2}l^2} dS ^2-\frac{\pi ^2Q}{S^{3/2}}dS dQ-\frac{3S^{1/2}}{l^3}dS dl+ \frac{\pi^2}{S^{1/2}}dQ^2
+ \frac{3S^{3/2}}{l^4   } dl^2\right),
\ee

\be
g^{II} = \frac{\beta_M M}{\pi^{3/2}}\left(-\frac{3S^2-\pi S l^2+3\pi^2 Q^2l^2}{8 S^{5/2}l^2} dS ^2+\frac{\pi^2}{S^{1/2}}dQ^2
+ \frac{3S^{3/2}}{l^4} dl^2\right),
\ee
\begin{multline}
    g^{III}=\frac{\beta_M M}{\pi^{3/2}}\Bigg(\frac{(3S^2+\pi S l^2-\pi^2Q^2l^2)(3S^2-\pi S l^2+3\pi^2 Q^2l^2)}{16\pi ^{3/2}S^3l^4} dS^2+\frac{\pi ^{5/2}Q^2}{S}dQ^2\\-\frac{\pi ^{1/2}Q(3S^2+\pi S l^2+\pi^2Q^2l^2)}{4S^2l^2}dS dQ-\frac{3(-S^2-\pi S l^2+\pi Q^2 l^2)}{4\pi^{3/2}l^5}dSdl-\frac{3S^3}{\pi^{3/2}l^6}dl^2 \Bigg).
\end{multline}

The general structure of the corresponding independent curvature scalars has been analyzed in Appendix \ref{km}, where we obtained in 
Eqs.(\ref{c11})--(\ref{c33}) the general conditions that relate the singularities of the three GTD metrics, which in this case can be written as 
\begin{align}
    I&: M_{,SS}M_{,ll}M_{,QQ}-\big(M_{,Sl}\big)^2M_{,QQ}-\big(M_{,SQ}\big)^2M_{,ll}=-\frac{3T}{2\pi l^4}=0, \label{RNcondI}
    \\
     II&: M_{,SS}M_{,QQ}M_{,ll}=\frac{3S}{\pi l^4}\big(3S^2-\pi S l^2 +3\pi^2 l^2Q^2\big)=0,\\
     III&: M_{,SS}=M_{,SQ}=M_{,Sl}=0, \quad \text{or}  \quad M_{,SQ}=M_{,QQ}=0,\nonumber \\ &\text{or} \quad M_{,Sl}=M_{,ll}=0,\quad \text{or} \quad M_{,QQ}=M_{,ll}=0. \label{RNcondIII}
\end{align}

\begin{figure}
\begin{minipage}[t]{0.47\linewidth}
 \centering
\hspace{1cm}
\includegraphics[width=1\linewidth]{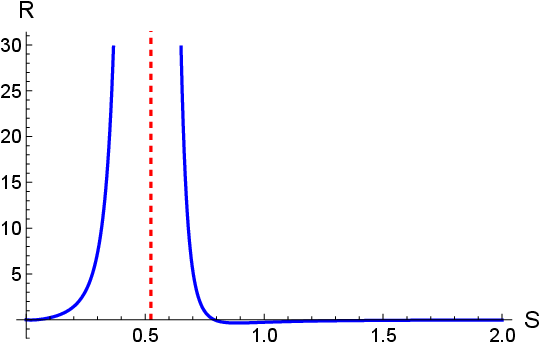}
 (a)\hspace{5cm}
\end{minipage}%
\hfill%
\begin{minipage}[t]{0.5\linewidth}
 \centering
\hspace{1cm}
\includegraphics[width=1\linewidth]{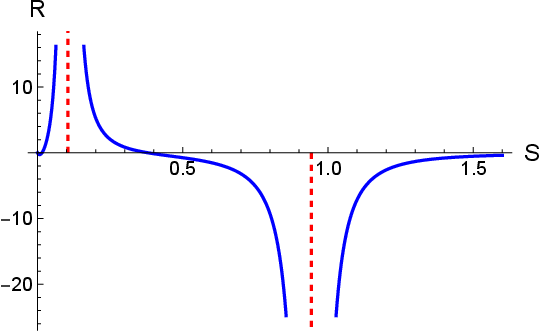}
(b)\hspace{5cm}
\end{minipage}
\begin{minipage}[t]{0.5\linewidth}
 \centering
\hspace{1cm}
\includegraphics[width=1\linewidth]{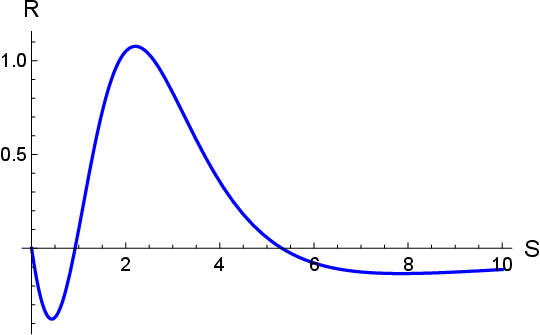}
 (c)\hspace{5cm}
\end{minipage}
    \caption{Behavior of the Ricci scalar $R^{II}$ of the RN-AdS black hole in terms of the entropy $S$ for different values of electric charge $Q$, radius of curvature $l=1$, and $\beta_M=1$. (a) $Q=l/6$. $R^{II}$ is singular at $S\approx 0.52$. (b) $Q=l/10$. $R^{II}$ has two singularties, one at $S\approx 0.10$ and other at $S\approx 0.94$.   (c) $Q=l$. The Ricci scalar is regular everywhere.
  \label{R2_RNAdS}}
\end{figure}

Condition $I$ is true only for the extremal case, i.e., $T=0$, and condition $III$ cannot be fulfilled in general. Instead, condition $II$ is fulfilled in general when $  S=(\pi l/6)\big( l \pm \sqrt{l^2-36Q^2}\big)$.
Thus, for $Q=l/6$, $R^{II}$ has only one singularity at $S=\pi l^2/3$. For values of $Q<l/6$, $R^{II}$ has two singularities  and for values of $Q>l/6$, $R^{II}$ is regular everywhere. In Fig. \ref{R2_RNAdS}, we illustrate the behavior of $R^{II}$, which in this case can be written as 
\be
    R^{II}=\frac{N^{II}}{24\beta_M\pi^{9/2}l^6S^{3/2} M^3\Big[3S^2-\pi S l^2 +3\pi^2 l^2Q^2\Big]^2}\label{rt2}
\ee
   \begin{multline}
        N^{II}=-8\pi^3Sl^4\Big[36S^6+27\pi S^5l^2+2\pi^2S^4l^2\big(7l^2-90Q^2\big)+2\pi^3 S^3l^4\big(2l^2-9Q^2\big)\\-\pi^4S^2Q^2l^4\big(108Q^2+59l^2\big)-9\pi^5SQ^4l^6+18\pi^6l^6Q^6\Big].
    \end{multline}

Additionally, using the results presented in   Appendix \ref{km} and the response functions of the system, we can write the singularity conditions Eqs.(\ref{RNcondI})--(\ref{RNcondIII}) as follows    
\begin{align}
    I&: \frac{T}{C_{L\phi}\kappa_{S\phi}\kappa_{Sl}}=0 ,\label{uy2}\\
    II&: \frac{T\kappa_{S\phi}\kappa_{Sl} }{C_{lQ}}=0,\label{qa}\\
    III&: \frac{1}{C_{lQ}}=\frac{1}{\alpha_{SQ}}=\frac{1}{\alpha_{Sl}}=0,\quad \text{or} \quad
    \frac{1}{\alpha_{SQ}}=\frac{1}{\kappa_{S,\phi}}=0,\label{rre3}\\ \nonumber
    &\text{or} \quad \frac{1}{\alpha_{S l}}=\frac{1}{\kappa_{S,l}}=0,\quad\text{or} \quad  \frac{1}{\kappa_{S\phi}}=\frac{1}{\kappa_{Sl}}=0;
\end{align}
\\
 where $\phi$ is the electrical potential wich is the dual variable of $Q$, i.e., $\phi= \partial M/\partial Q$. In turn, the response functions are given explicity as
\begin{align}
C_{lQ}&=T\Big(\frac{\partial S}{\partial T}\Big)_{lQ}=\frac{6 S^3+2\pi l^2 S (S-\pi Q^2)}{3S^2-\pi S l^2 +3\pi^2 l^2Q^2},\\
C_{L\phi}&=T\Big(\frac{\partial S}{\partial T}\Big)_{L\phi}=-2S,\\
\kappa_{Sl}&=\Big(\frac{\partial Q}{\partial \phi}\Big)_{Sl}=\frac{S^{1/2}}{\pi^{1/2}},\\
\kappa_{S\phi}&=\Big(\frac{\partial l}{\partial L}\Big)_{S\phi}=\frac{l^4\pi^{3/2}}{3S^{3/2}},\\
\alpha_{Sl}&=\Big(\frac{\partial Q}{\partial T}\Big)_{Sl}=\frac{-2S^{3/2}}{\pi^{1/2}Q},\\
\alpha_{SQ}&=\Big(\frac{\partial l}{\partial T}\Big)_{SQ}=\frac{-2l^3\pi^{3/2}}{3S^{1/2}}.\\
\end{align}
 

Therefore, from Eqs.(\ref{uy2})--(\ref{rre3}) it is clear that the singularity $II$ coincides with the divergences of $C_{lQ}$. Notice that  for the  RN-AdS black hole the heat capacity $C_{L\phi}$, all compressibility  parameters, and thermal coefficients are regular. Consequently, like the uncharged case, the condition $I$ is only  true for $T=0$, and the condition $III$ is not fulfilled in general.

\begin{figure}[H]
    \centering
    \includegraphics[width=0.8\linewidth]{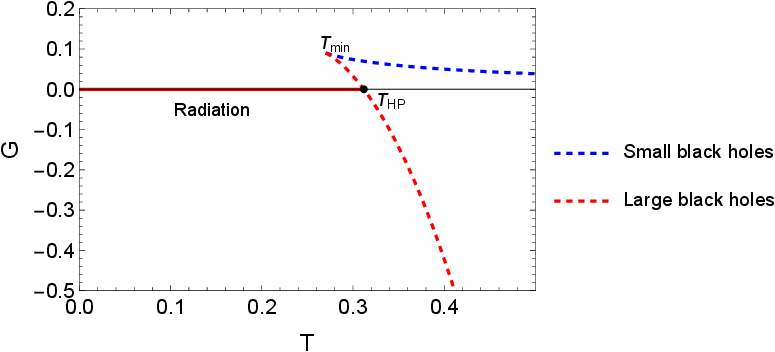}
    \caption{Gibbs free energy of the RN-AdS black hole for fixed $\phi=0.2$, and $l=1$.}
        \label{gibbs_chargedfigu}
\end{figure}

Next, we aim to study the phase transition structure of the charged AdS black hole using the Gibbs free energy of the system. Notably, a charged black hole in the canonical ensemble (fixed value of $Q$), due to the conservation of charge, will not  undergo a phase transition to thermal vacuum, which is electrically neutral \cite{WangRuppeiner}. Therefore, we wil use  the grand canonical ensemble (fixed electric potential $\phi$) to examine the Hawking-Page transition of the RN-AdS black hole.  First, we need to express the Hawking temperature Eq.(\ref{temp_charged}) in terms of the thermodynamic quantities $r_h$, $l$, and $\phi$, namely
\begin{equation}
\label{tem_grandcano}
   T=\frac{3r_h^2+l^2(1-\phi^2)}{4 \pi l^2 r_h}.
\end{equation}
Considering the mass as enthalpy, the Gibbs free energy in the grand canonical ensemble has the form 
\begin{equation}
\label{gibbs_charged}
    G(T,l,\phi)=M-TS-\phi Q=\frac{r_h(1-\phi^2)-r_h^3/l^2}{4},
\end{equation}
where $r_h$ is understood as a function of $T, l$, and $\phi$. 
Solving  the temperature equation (\ref{tem_grandcano}) with respect to $r_h$, we obtain 
\begin{equation}
r_h=\frac{l}{3}\Big[2\pi l T\pm \sqrt{4 \pi^2l^2T^2+3(\phi^2-1)}\Big].
\end{equation}
Accordingly, the Gibbs energy has two branches (see Fig. \ref{gibbs_chargedfigu}) and is defined for temperatures greater than
\begin{equation}
    T_{min}=\frac{\sqrt{3(1-\phi^2)}}{2 \pi l}.\label{tmin_charged}
\end{equation}

In the grand canonical ensemble, we allow that the particle number of the thermal gas varies with temperature, and,  as was pointed out in \cite{hawking1983thermodynamics}, the Gibbs free
energy of thermal AdS background is zero. Thus, from Eq. (\ref{gibbs_charged}), we obtain the vanishing point of the Gibbs free energy  
\begin{equation}
  r_h(1-\phi^2)-r_h^3/l^2=0,
\end{equation}
which gives the Hawking-Page temperature\footnote{Note that  Eqs. (\ref{tem_grandcano})--(\ref{THp_charged}) for $\phi=0$  reduce to the expressions obtained in the Schwarzschild case.} as
\begin{equation}
    \label{THp_charged}
    T_{HP}=\frac{\sqrt{(1-\phi^2)}}{ \pi l}.
\end{equation}
For temperature values lower than this threshold, thermal radiation dominates the background. As the temperature rises, there is a chance for thermal radiation to lead to the creation of a large black hole, as indicated by the segment between $T_{min}$ and $T_{HP}$ in Fig. \ref{gibbs_chargedfigu}. The Hawking-Page transition occurs at $T_{HP}$, above which the formation of a  stable large black hole is more viable. 


\section{RN-AdS shadows from GTD}
\label{sec:shgtd}
In this section, we analyze the shadow of the RN-AdS black hole using results obtained on the previous section  from  the GTD formalism. Specifically, we examine the behavior of the curvature scalar $R^{II}$. First, we investigate the divergences of $R^{II}$ in relation to the horizon radius $r_{h}$. Utilizing Eq.\eqref{rt2}, the fundamental equation \eqref{fundame2}, and the entropy expression $S = \pi r_h^2$, we find the explicit expression for $R^{II} = R^{II}(r_h)$. Next, we demonstrate that the shadow radius $r_{sh}$ acts as an observable parameter reflecting the thermodynamic behavior of the horizon radius $r_{h}$ in the scalar $R^{II}$. This is achieved by analyzing $R^{II}(r_{sh})$, which is obtained by linking Eq.\eqref{shadowofrh} for the shadow radius $r_{sh}(r_h)$ with $R^{II}(r_h)$. Additionally, to further illustrate the system's behavior in the thermodynamic phase space, we present the variation of the scalar $R^{II}$ across the shadow's profile using Eqs.\eqref{Projection1} and \eqref{Projection2}.
  \begin{figure}[H]
\begin{minipage}[t]{0.435\linewidth}
 \centering
\hspace{1cm}
\includegraphics[width=1\linewidth]{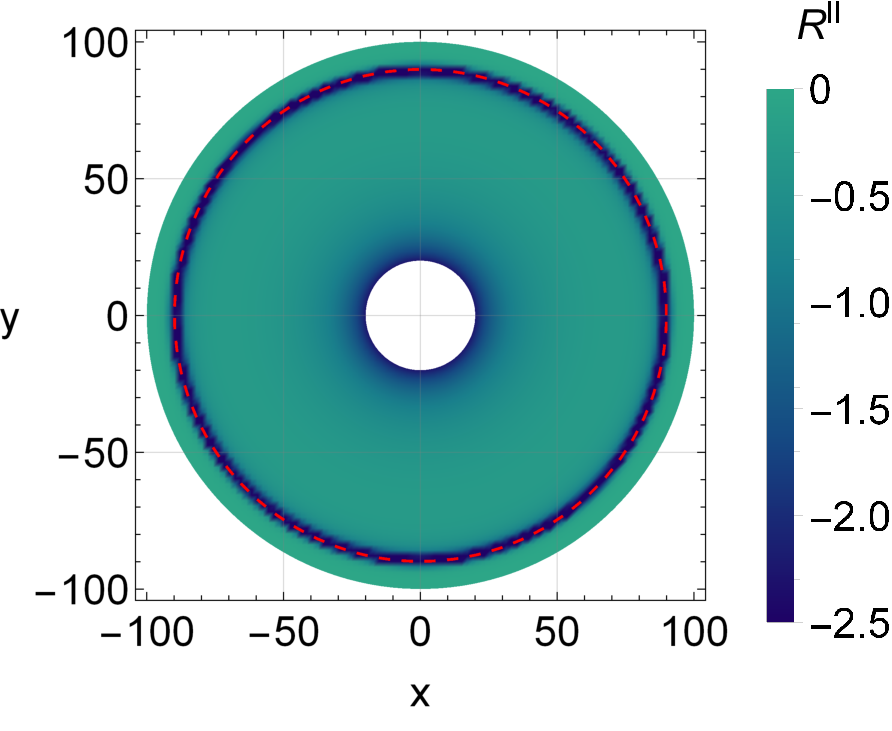}
 (a)\hspace{5cm}
\end{minipage}%
\hfill%
\begin{minipage}[t]{0.515\linewidth}
 \centering
\hspace{1cm}
\includegraphics[width=1\linewidth]{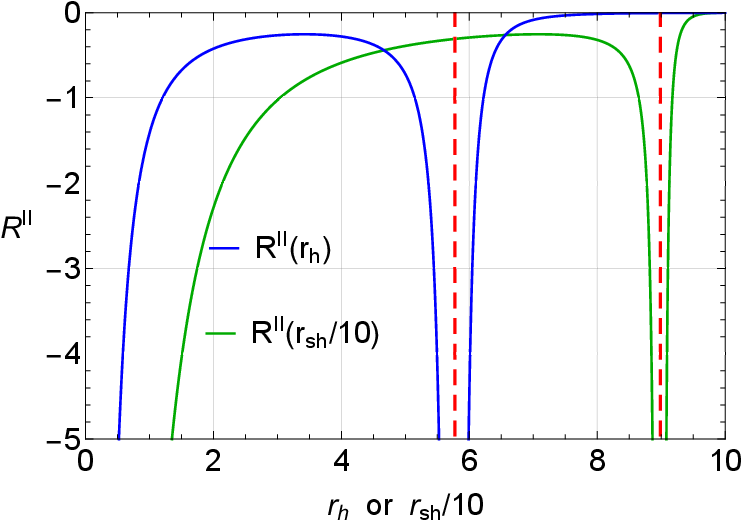}
(b)\hspace{5cm}
\end{minipage}
    \caption{(a) The variation of the scalar $R^{II}(r_{sh})$ across the shadow's cast profile. (b) Scalar $R^{II}$ as a function of the horizon radius $r_{h}$ and the scaled shadow radius $r_{sh}/10$, for a Schwarzschild-AdS black hole. (with $\beta_M=10$, $r_O=100$ and $l=10$). The red dotted lines indicate the divergences of $R^{II}$ (except $r_{h}=0$). \label{Figure6}}
\end{figure}

Starting with the case of a Schwarzschild-AdS black hole, in Fig. \ref{Figure6}, we illustrate the variation of the scalar $R^{II}$ across the shadow's cast profile and as a function of $r_{h}$ and $r_{sh}/10$ (with $r_O=100$ and $l=10$). Here, the shadow's cast profile and $R^{II}$ confirm that in this case, the black hole behaves like a van der Waals system, undergoing a second-order phase transition at $T = T_c$. This transition is reflected by a divergence at $r_{h} \approx 5.7735$ and $r_{sh} \approx 89.8546$. $R^{II}$ shows consistent results for both $r_{h}$ and $r_{sh}$, and its behavior matches that of $C_l$ in Fig. \ref{Figure1}. In Fig.  \ref{Figure6}, a divergence of $R^{II}$ is also observed when $r_{h}=0$. However, this is discarded in the analysis since $r_{h}=0$ is equivalent to the non-existence of the black hole. Strictly speaking, it should be $r_{h}>0$. Therefore, the shadow's cast profile has been evaluated in the range of $20<r_{sh}<100$.

\begin{figure}
\begin{minipage}[t]{0.435\linewidth}
 \centering
\hspace{1cm}
\includegraphics[width=1\linewidth]{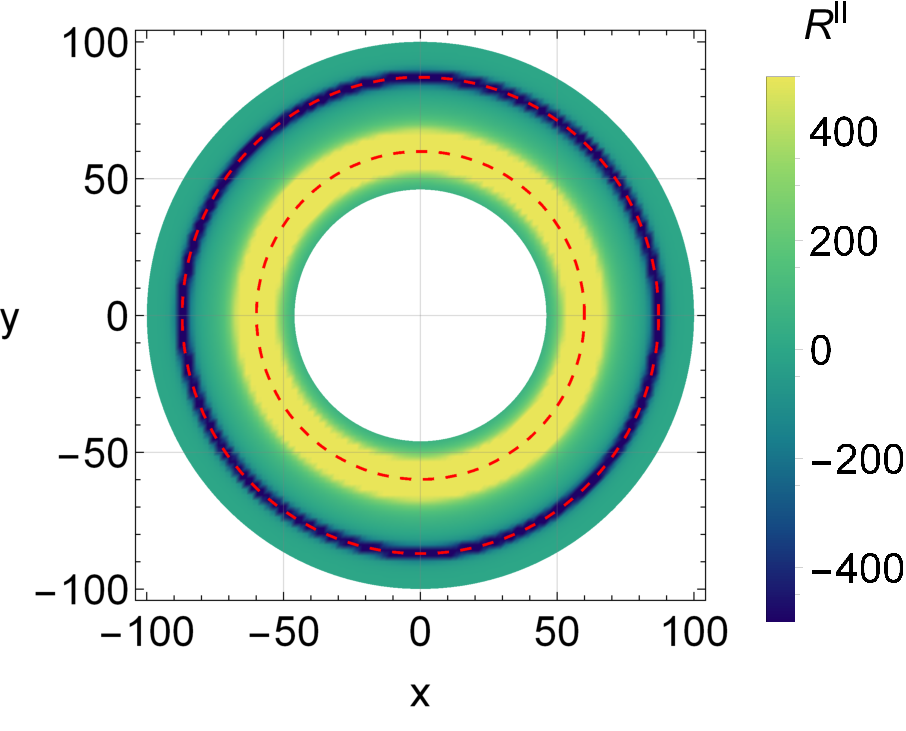}
 (a)\hspace{5cm}
\end{minipage}%
\hfill%
\begin{minipage}[t]{0.515\linewidth}
 \centering
\hspace{1cm}
\includegraphics[width=1\linewidth]{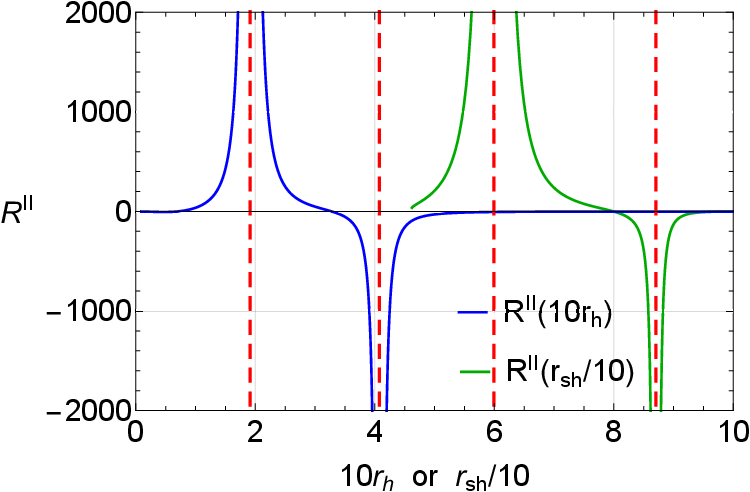}
(b)\hspace{5cm}
\end{minipage}
    \caption{(a) The variation of the scalar $R^{II}(r_{sh})$ across the shadow's cast profile of a RN-AdS black hole. (b) Scalar $R^{II}$ as a function of the scaled horizon radius $10r_{h}$ and the scaled shadow radius $r_{sh}/10$. (with $\beta_M=10$, $r_O=100$, $Q=0.1$ and $l=1.3l_{c}$).The red dotted lines indicate the divergences of $R^{II}$. \label{Figure7}}
\end{figure}

\begin{figure}[H]
\begin{minipage}[t]{0.435\linewidth}
 \centering
\hspace{1cm}
\includegraphics[width=1\linewidth]{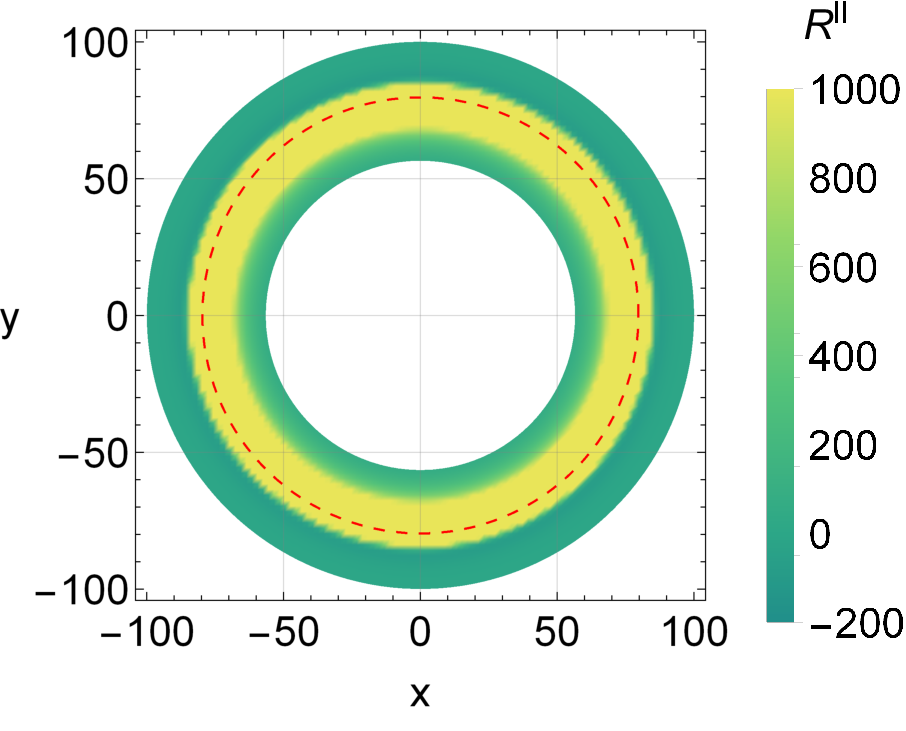}
 (a)\hspace{5cm}
\end{minipage}%
\hfill%
\begin{minipage}[t]{0.515\linewidth}
 \centering
\hspace{1cm}
\includegraphics[width=1\linewidth]{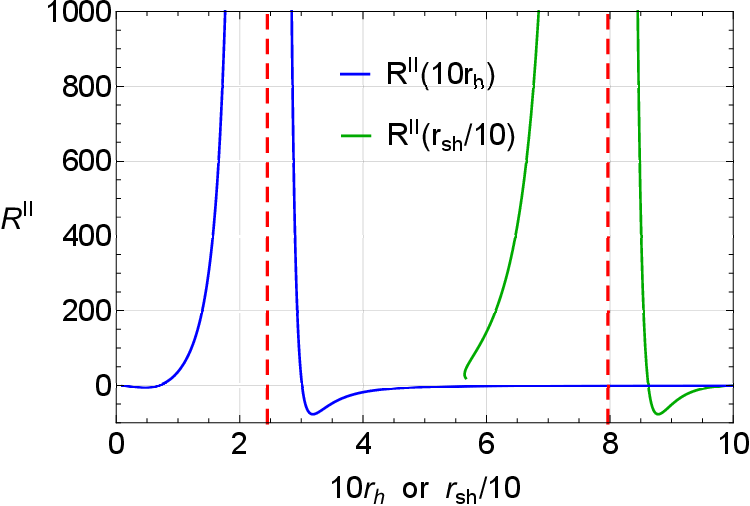}
(b)\hspace{5cm}
\end{minipage}
    \caption{(a) The variation of the scalar $R^{II}(r_{sh})$ across the shadow's cast profile of a RN-AdS black hole. (b) Scalar $R^{II}$ as a function of the scaled horizon radius $10r_{h}$ and the scaled shadow radius $r_{sh}/10$. (with $\beta_M=10$, $r_O=100$, $Q=0.1$ and $l=l_{c}$). The red dotted lines indicate the divergences of $R^{II}$. \label{Figure8}}
\end{figure}

\begin{figure}
\begin{minipage}[t]{0.435\linewidth}
 \centering
\hspace{1cm}
\includegraphics[width=1\linewidth]{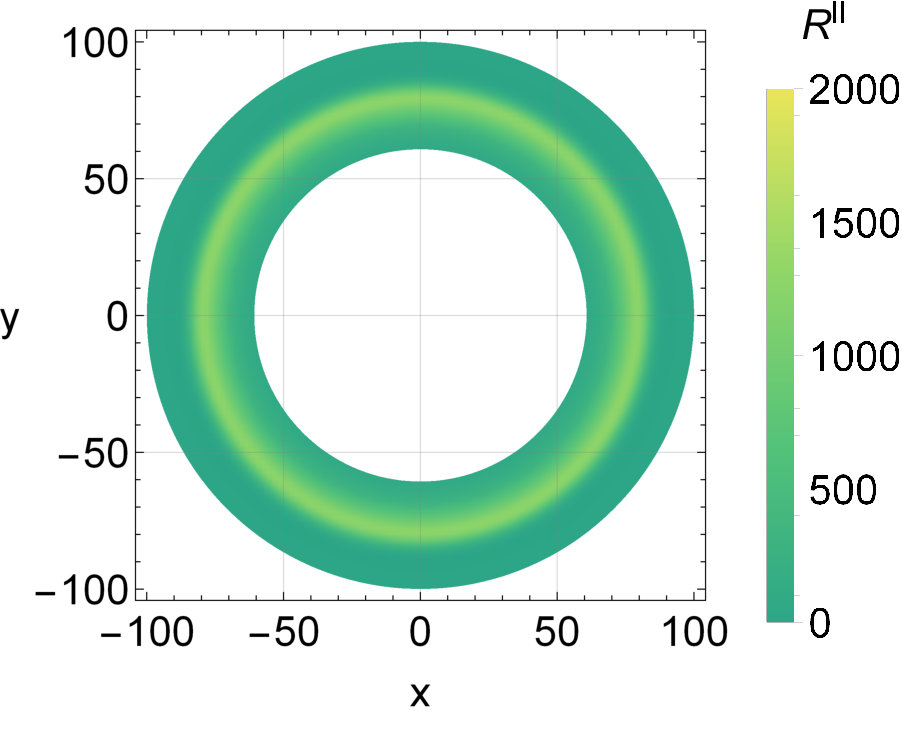}
 (a)\hspace{5cm}
\end{minipage}%
\hfill%
\begin{minipage}[t]{0.515\linewidth}
 \centering
\hspace{1cm}
\includegraphics[width=1\linewidth]{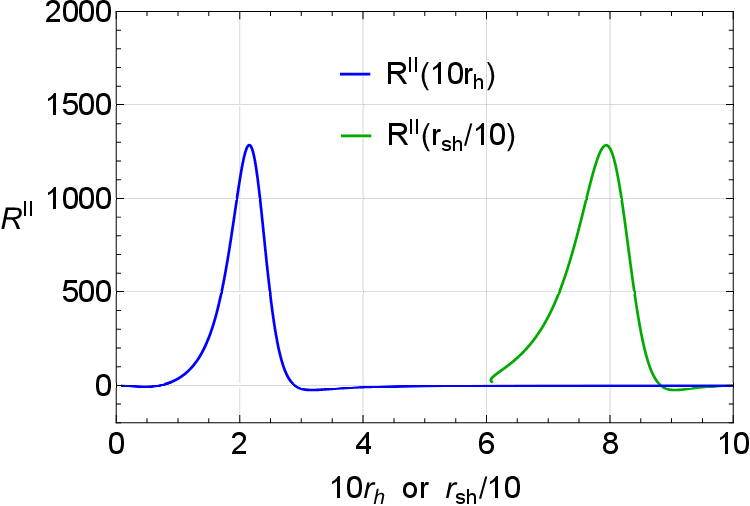}
(b)\hspace{5cm}
\end{minipage}
    \caption{(a) The variation of the scalar $R^{II}(r_{sh})$ across the shadow's cast profile of a RN-AdS black hole. (b) Scalar $R^{II}$ as a function of the scaled horizon radius $10r_{h}$ and the scaled shadow radius $r_{sh}/10$. (with $\beta_M=10$, $r_O=100$, $Q=0.1$ and $l=0.9l_{c}$). \label{Figure9}}
\end{figure}

In the case of RN-AdS black holes, we present the variation of the scalar $R^{II}$ across the shadow's cast profile and as a function of $r_{h}$ and $r_{sh}/10$ (with $r_O=100$ and $Q=0.1$) in Figs.  \ref{Figure7}, \ref{Figure8}, and \ref{Figure9}. These figures illustrate different values of $l$. Considering the possible values where $r_{sh}$ reflects a meaningful thermodynamic behavior, as discussed in our previous analysis of Fig.\ref{Figure3}, we use Eq.\eqref{extremeT} in \eqref{shadowofrh} to determine $r_{min}$. For $r_O = 100$, the shadow profiles have been evaluated in the ranges $46.1972 < r_{sh} < 100$, $56.4483 < r_{sh} < 100$, and $60.7065 < r_{sh} < 100$ in Figs. \ref{Figure7}, \ref{Figure8}, and \ref{Figure9}, corresponding to the cases $l > l_c$, $l = l_c$, and $l < l_c$, respectively.

In Fig. \ref{Figure7}, for the case where $l>l_c$, $R^{II}$ exhibits two divergences at $r_{h}=0.1913$ and $r_{h}=0.4077$ for the horizon radius, and $r_{sh}=59.9239$ and $r_{sh}=87.0504$ for the shadow radius. These discontinuities in $R^{II}$ align with the divergences observed in $C_{lQ}$ in Fig. \ref{Figure4}. The non-monotonic behavior of the temperature, characterized by a local maximum and minimum, indicates the presence of two second-order phase transitions because they correspond to  divergences of the heat capacity $C_{lQ}$.
 Notably, these points coincide with a change in sign of $R^{II}$, transitioning from positive to negative at the first divergence and remaining  negative at the second divergence. 
RN-AdS black holes with $r_{sh}$ smaller than the first divergence of $R^{II}$ are stable, while the region after the second divergence in heat capacity relates to thermally stable black holes, as indicated by the positive slope of $R^{II}$ and $T$. Conversely, black holes within the intermediate range of $r_{sh}$ are thermodynamically unstable, as evidenced by the negative slope in $R^{II}$ and $T$ 
\cite{ZhangGuo, Luo, GuoLi, Belhaj, WangRuppeiner, Li, Kumar, Du, Zheng, He}.

In Fig.  \ref{Figure8}, corresponding to the case where $l=l_c$, $R^{II}$ exhibits a divergence at $r_{h}=0.2449$ and $r_{sh}=79.6705$. These discontinuities align with the divergences observed in $C_{lQ}$ on the left-hand side of Figure \ref{Figure5}. For this case, where $l=l_c$, the scalar $R^{II}$ demonstrates behavior reminiscent of a van der Waals system. The smallest and largest black holes merge into one, resulting in a thermodynamically unstable black hole. Therefore, the system undergoes a second-order phase transition, which is reflected in an inflection point in $T$ and a divergence in $C_{lQ}$ 
\cite{ZhangGuo, Luo, GuoLi, Belhaj, WangRuppeiner, Li, Kumar, Du, Zheng, He}. In Figure \ref{Figure9}, corresponding to the case where $l<l_c$, $T$ increases monotonically, as explicitly shown by the behavior of $C_{lQ}$ on the right-hand side of Fig. \ref{Figure5}. In this case, there are no divergences in $R^{II}$ or $C_{lQ}$, indicating that the black hole is in the supercritical phase
\cite{ZhangGuo, Luo, GuoLi, Belhaj, WangRuppeiner, Li, Kumar, Du, Zheng, He}.

\section{On the microstructure of the RN-AdS black hole}
\label{sec:mic}

 \begin{figure}
\begin{minipage}[t]{0.47\linewidth}
 \centering
\hspace{1cm}
\includegraphics[width=1\linewidth]{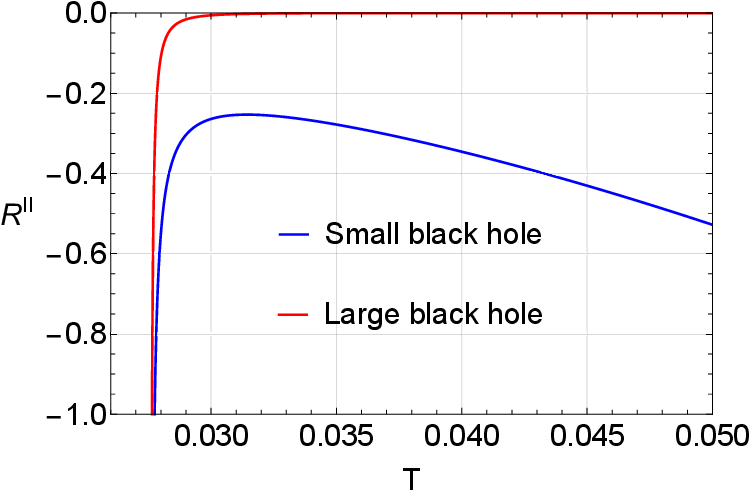}
 (a)\hspace{5cm}
\end{minipage}%
\hfill%
\begin{minipage}[t]{0.49\linewidth}
 \centering
\hspace{1cm}
\includegraphics[width=1\linewidth]{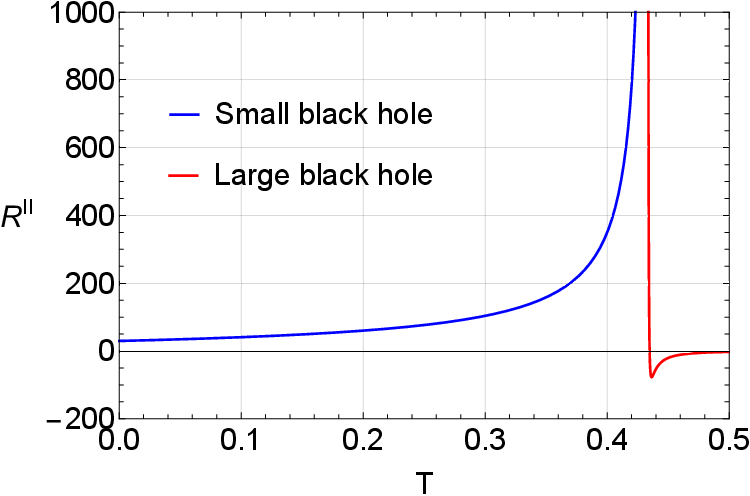}
(b)\hspace{5cm}
\end{minipage}
    \caption{The behavior of the scalar $R^{II}$ against the temperature $T$. (a) For a Schwarzschild AdS black hole. (with $l = 10$ and $\beta_M=10$ ). (b) For a RN AdS black hole. (with $Q=0.1$, $l = l_c$  and $\beta_M=10$ ). \label{Figure10}}
\end{figure}
 \begin{figure}
\begin{minipage}[t]{0.47\linewidth}
 \centering
\hspace{1cm}
\includegraphics[width=1\linewidth]{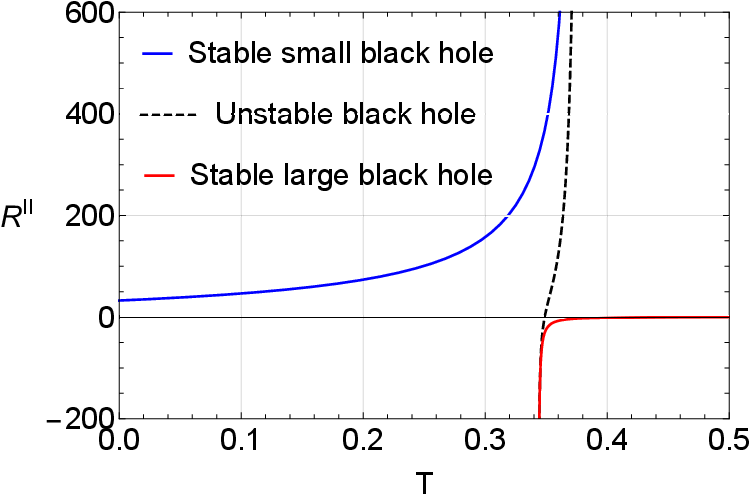}
 (a)\hspace{5cm}
\end{minipage}%
\hfill%
\begin{minipage}[t]{0.49\linewidth}
 \centering
\hspace{1cm}
\includegraphics[width=1\linewidth]{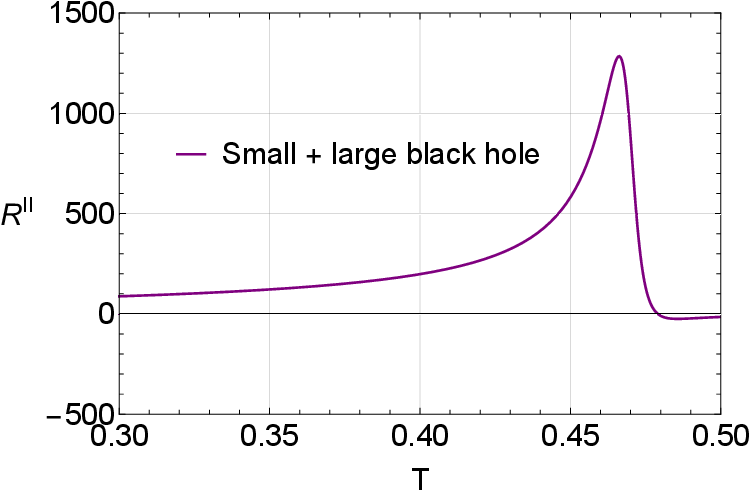}
(b)\hspace{5cm}
\end{minipage}
    \caption{The behavior of the scalar $R^{II}$ against the temperature $T$ for a RN AdS black hole. (a) With $Q=0.1$, $l = 1.3 l_c$  and $\beta_M=10$. (b) With $Q=0.1$, $l = 0.9 l_c$  and $\beta_M=10$. \label{Figure11}}
\end{figure}

In Figs. \ref{Figure10} and \ref{Figure11}, we show the behavior of the curvature scalar $R^{II}$ in terms of the temperature $T$, highlighting the microstructure for the corresponding stabilities that define the phase transitions of small and large black holes. We can see that in the case of a Schwarzschild-AdS black hole, $R^{II}$ always remains negative for both small unstable and large stable black holes, indicating a dominance of attractive interaction in the microscopic system. This is in agreement with the results already obtained using Ruppeiner geometry in \cite{ref14}.
The microstructure of RN-AdS black holes  revealed by thermodynamic geometry  exhibits a very different behavior compared to our results given by the GTD formalism. In various studies \cite{ref15, ref16, ref18, ref19, WangRuppeiner, ref20, ref21}, it has been observed that the predominant microstructure, according to Ruppeiner geometry, shows repulsive interactions only in the case of small black holes, reminiscent of anyon gas behavior. Conversely, for high-temperature small black holes, the interactions tend to be attractive. Furthermore, for large black holes, Ruppeiner geometry typically indicates exclusively attractive interactions, although a study \cite{ref17} noted an attractive interaction domain specifically within the regime of low-temperature large black holes. As we will detail below, the GTD formalism reveals that the microstructure of RN-AdS black holes is repulsive only for small black holes, while for large black holes, it exhibits attractive interactions, similar to anyon gas behavior. In some cases, for low-temperature large black holes the interactions tend to be repulsive. This highlights a novel perspective from GTD on the true effective behavior of the microstructure of black holes.

For RN AdS-black holes, we analyze three cases with $l > l_c$, $l = l_c$, and $l < l_c$. Using Eq.\eqref{Tminmax} for the local minimum and maximum temperature values, we can verify that these coincide with the divergences of $R^{II}$. As seen in Fig.  \ref{Figure10}, when $l = l_c$, for small black holes, $R^{II}$ is always positive, thus the microstructure interaction is repulsive. However, for large black holes, near the critical temperature point, repulsive interaction dominates. At $T=0.4348$, $R^{II}=0$, indicating no effective interaction between the microscopic molecules. For higher temperatures, attractive interaction persists in stable large black holes. On the other hand, in Fig.  \ref{Figure11}, in the case where $l > l_c$, for small black holes, $R^{II}$ is always positive, indicating repulsive microstructure interaction. For large black holes, attractive interaction dominates as $R^{II}$ remains negative. At $T=0.3488$, $R^{II}=0$, but this occurs in the unstable black hole state. Additionally, in the case where $l < l_c$, the supercritical black hole state is observed, where small and large black holes cannot be distinguished. In this case, $R^{II}$ is positive, with repulsive microstructure interaction, up to $T=0.4791$ where $R^{II}=0$. After this, for higher temperatures, $R^{II}$ remains negative, with predominantly attractive interaction. In general, these results on microscopic behavior are different from those obtained using thermodynamic geometry. The GTD approach indicates that the microstructure of RN-AdS black holes exhibits repulsive interactions predominantly in small black holes. Otherwise, larger black holes show attractive interactions akin to the behavior seen in anyon gases. At low temperatures, large black holes can also display repulsive interactions. This presents a new viewpoint from the GTD on the microstructure in RN-AdS black holes.


\section{Conclusions}
 \label{sec:con}

The main purpose of the present work was to investigate the thermodynamic properties and microstructure of the RN-AdS black hole, using the formalism of GTD. First, we established for this black hole the general relationships between shadow properties and thermodynamic quantities. Then, we analyzed for both the Schwarzschild-AdS and the RN-AdS black hole the corresponding equilibrium space by using the GTD. To guarantee the consistency of the GTD, it was required to interpreter the curvature radius as thermodynamic variable, and hence describe the black hole as a quasi-homogeneous system. We found that the radius of curvature of the RN-AdS black hole affects strongly the structure and disposition of the curvature singularities of the equilibrium space, which were shown to be in strict correspondence with the phase transition structure, following from the behavior of the black hole response functions. In general, it was shown that the RN-AdS black hole can have zero, one, or two second order phase transitions, depending on the value of the radius of curvature. Furthermore, we analyzed the Gibbs free energy, and confirmed that the phase transition predicted by GTD corresponds to a small-large AdS black hole phase transition, which is similar to the case of the van der Walls fluid. We also found that for the charged scenario, we require the grand-canonical ensemble to study the Hawking-Page phase transition. A result that underscores the dependence of the black hole phase transition structure on the underlying statistical ensemble. Therefore, due to the lack of a fully consistent statistical description of black holes, our GTD characterization of the phase transition structure in terms of the curvature of the equilibrium space allows us to perform an invariant analysis of the correspondence between shadows and phase transitions.

Through the analysis of phase transitions in black holes within the context of shadows, we can draw several significant conclusions. Firstly, the shadow radius can effectively replace the event horizon radius in capturing the phase transition process. Moreover, the critical thermodynamic behavior and the microstructure of AdS black holes can be revealed by examining the shadow radius. This suggests that the shadow radius has the potential to serve as a valuable indicator for the phase structure of black holes. In our study, we have successfully applied GTD to the realm of shadow thermodynamics. By doing so, we have demonstrated that this geometrical approach is a powerful and novel tool for analyzing the phase transitions and the microstructure of black holes using their shadows. 

We expect to utilize GTD to explore alternative gravity theories within the framework of black hole shadows. Future work will be to apply this GTD approach to the shadows of rotating black holes as well. Furthermore, we have derived parameter restrictions that, if applied analogously to alternative gravity theories, could provide valuable insights for confirming or ruling out different aspects of these theories. This approach would involve utilizing real data, such as the observations obtained by the Event Horizon Telescope collaboration, which have provided detailed information on the shadows of M87* and Sgr A*.

\section*{Acknowledgments}

This work was partially supported by Conahcyt-Mexico, Grant No. A1-S-31269. 
JML and CRF acknowledge support from Conahcyt-Mexico.


\appendix

 \section{Two-dimensional geometrothermodynamics}
\label{app:2dof}

Consider the case of a system with two thermodynamic degrees of freedom ($n =2$). From Eqs.(\ref{g1})-(\ref{g3}), we obtain in this case \cite{quevedo2023unified}
\begin{equation}
    g^I=\beta_\Phi \Phi \Big[\Phi_{,11} (dE^1)^2+\Phi_{,22} (dE^2)^2+2\Phi_{,12} dE^1dE^2\Big],\label{2g11}
\end{equation}
\be
   g^{II}=\beta_\Phi \Phi \Big[-\Phi_{,11} (dE^1)^2+\Phi_{,22} (dE^2)^2\Big], \label{g222}
\end{equation}
\bea 
g^{III}=  & \beta_1 \big(E^1 \Phi_{,1} \big)^{2k+1}\Phi_{,11}(dE^1)^2+\beta_2 \big(E^2\Phi_{,2} \big)^{2k+1}\Phi_{,22}(dE^2)^2 \nonumber \\
& +\Big[\beta_1 \big(E^1 \Phi_{,1} \big)^{2k+1}+\beta_2 \big(E^2 \Phi_{,2} \big)^{2k+1}\Big]\Phi_{,12}dE^1dE^2,\label{g233}
\eea 
  where $\Phi_{,a}=\partial \Phi/\partial E^a$. To investigate the singularity structure of the above metrics, we compute the corresponding scalar curvature\footnote{For a two-dimensional space, the  scalar curvature completely determines the Riemann tensor.}. In doing this, we demand that the singularities of $g^{III}$ are related to those of $g^I$ and $g^{II}$ so that all the metrics can be used to describe the same system. It then follows that this condition fixes the value of the integer $k$ entering the metric $g^{III}$ as $k =0$. Then, a straight-forward computation leads to the following Ricci scalars \cite{quevedo2023unified}

\begin{equation}
   R^I=\frac{N^I}{D^I}, \quad D^I= 2 \beta_\Phi \Phi^3 \Big [\Phi_{,11}\Phi_{,22}-\big (\Phi_{,12}\big)^2  \Big]^2 ,\label{d11}
\end{equation}
\begin{equation}
     R^{II}=\frac{N^{II}}{D^{II}}\quad D^{II}= 2 \beta_\Phi \Phi^3\Big ( \Phi_{,11}\Phi_{,22} \Big)^2,\label{d22}
\end{equation}
\begin{equation}
      R^{III}=\frac{N^{III}}{D^{III}}\quad D^{III}=\Big [ \beta_\Phi^2 \Phi^2\big (\Phi_{,12}\big)^2 -4\beta_1\beta_2E^1E^2\Phi_{,1}\Phi_{,2}\Phi_{,11}\Phi_{,22}\Big]^2. \label{d33}
\end{equation}

  The singularities of the equilibrium space metrics are determined by the zeros of the functions $D^I$, $D^{II}$ and $D^{III}$. For $D^I$, the only non-trivial zero,  $\Phi_{,11}\Phi_{,22}-\big (\Phi_{,12}\big)^2=0$, represents exactly the breakdown of the stability condition of a thermodynamic system with two degrees of freedom \cite{callen1998thermodynamics}. This means that the singularity of $R^I$ represents in general a phase transition. The singularities of  $R^{II}$, $\Phi_{,11}\Phi_{,22}=0$, are also related to phase transitions because they correspond to divergences of the response functions. Indeed, response functions usually represent the dependence of the extensive $E ^a$ variables in terms of the intensive variables $I_b$, i.e., $\frac{\partial E^a}{ \partial I_b}$. Then, we see that 
  \be
  \frac{\partial E^a }{\partial I_b} = 
  \left(\frac{\partial I_b}{\partial E^a}\right)^{-1} = \frac{1}{\Phi_{,ab}} \ .
 \ee
  So, we see that the singularities of $R^{II}$  coincide with the divergences of the response functions $\partial E^1 /\partial I_1 $ and $\partial E^2 /\partial I_2 $. As for the zeros of $D^{III}$, i.e., $\beta_\Phi^2 \Phi^2\big (\Phi_{,12}\big)^2 -4\beta_1\beta_2E^1E^2\Phi_{,1}\Phi_{,2}\Phi_{,11}\Phi_{,22}=0$, the situation is different because, in general, they cannot be associated directly with divergences of response functions. Therefore, in order for the three GTD metrics to be consistent when applied to the same system, we demand the three sets of singularities be related in the following sense. Let $R^I$ be singular, i.e., $(\Phi_{,12})^2 = \Phi_{,11} \Phi_{,22}$. Then, $D^{III}$ becomes 
  \be 
  D^{III} = [(\Phi_{,12})^2 ( \beta_\Phi^2 \Phi ^2 - 4 \beta_1 \beta_2 E^1 E^2 \Phi_{,1}\Phi_{,2})]^2,
  \ee
an expression which is zero for $\Phi_{,12}=0$ and or $\beta_\Phi \Phi ^2 - 4 \beta_1 \beta_2 E^1 E^2 \Phi_{,1}\Phi_{,2}=0$. 
This last condition is not satisfied in general because it fixes completely the fundamental equation as $\Phi(E^1,E^2) = \left(c_1\ln E^1 + c2\right)(c_3/c_1\ln E^2 +c_4)$, where the $c$'s are real constants. We conclude that in this case the only allowed zero is $\Phi_{,12}=0$. Now suppose that $R^{II}$ is singular, i.e., $\Phi_{,11}\Phi_{,22} =0$. Then, $D^{III} = (\beta_\Phi \Phi \Phi_{,12})^4$, for which the only non-trivial zero is $\Phi_{,12}=0$. We see that in both cases the compatibility of the singularities implies that the only allowed solution is $\Phi_{,12}=0$, which corresponds to a divergence of the response function $\partial E^2/\partial I_1$.

We conclude that if we demand that the singularities of $g^{III}$ be compatible with those of $g^I$ and $g^{II}$, all the singularities are determined by the zeros of the second-order derivatives of $\Phi$, namely, \cite{quevedo2023unified}
    \begin{align}
      I&:\Phi_{,11} \Phi_{,22}-\big (\Phi_{,12}\big)^2=0,\label{c1}\\
       II&:\Phi_{,11} \Phi_{,22}=0,\\
        III&: \Phi_{,12}=0,\label{c3}\ 
\end{align}
conditions that are known to indicate the presence of phase transitions.   
  

\section{Three- and higher-dimensional  geometrothermodynamics}
\label{km}

We will consider now the case of a system with three thermodynamic degrees of freedom $(n =3)$. Then, the fundamental equation reads $\Phi=(E^1,E^2,E^3)$. From Eqs.(\ref{g1})-(\ref{g3}), we obtain in this case
\begin{equation}
    g^I=\beta_\Phi \Phi \Big[\Phi_{,11} (dE^1)^2+\Phi_{,22} (dE^2)^2+\Phi_{,33} (dE^3)^2+2\Big(\Phi_{,12} dE^1dE^2+\Phi_{,13} dE^1dE^3+\Phi_{,23} dE^2dE^3\Big)\Big],\label{g11}
\end{equation}
\begin{equation}
     g^{II}=\beta_\Phi \Phi \Big[-\Phi_{,11} (dE^1)^2+\Phi_{,22} (dE^2)^2+\Phi_{,33} (dE^3)^2+2 \Phi_{,23} dE^2dE^3\Big], \label{g22}
\end{equation}
\begin{multline}
      g^{III}=\beta_1 \big(E^1 \Phi_{,1} \big)^{2k+1}\Phi_{,11}(dE^1)^2+\beta_2 \big(E^2\Phi_{,2} \big)^{2k+1}\Phi_{,22}(dE^2)^2 +\beta_3 \big(E^3 \Phi_{,3} \big)^{2k+1}\Phi_{,33}(dE^3)^2+\\\Big[\beta_1 \big(E^1 \Phi_{,1} \big)^{2k+1}+\beta_2 \big(E^2 \Phi_{,2} \big)^{2k+1}\Big]\Phi_{,12}dE^1dE^2+\Big[\beta_1 \big(E^1 \Phi_{,1} \big)^{2k+1}+\beta_3\big(E^3 \Phi_{,3} \big)^{2k+1}\Big]\Phi_{,13}dE^1dE^3+\\\Big[\beta_2 \big(E^2 \Phi_{,2} \big)^{2k+1}+\beta_3 \big(E^3 \Phi_{,3} \big)^{2k+1}\Big]\Phi_{,23}dE^2dE^3.\label{g33}
\end{multline}
We can write the above metrics in a compact way
\begin{equation}
    g= A_{11}(dE^1)^2+A_{22}(dE^2)^2+A_{33}(dE^3)^2+2\Big(A_{12}dE^1dE^2+A_{13}dE^1dE^3+A_{23}dE^2dE^3\Big),\label{metricaux}
\end{equation}
where the metric functions\footnote{The $ij$ index are just label, it does not mean derivative with respect $ij$ coordinate.} $A_{ij}$ are  functions of the extensive variables, i.e, $A_{ij}\equiv A_{ij}(E^1,E^2,E^3)$.
The computation of the scalar curvature for the above metric reads

\begin{equation}
     R=\frac{N(E^1,E^2,E^3)}{D(E^1,E^2,E^3)},\quad D\propto \Big [ A_{13}^2A_{22}-2 A_{12}A_{13}A_{23}+A_{12}^2A_{33}+A_{11}\Big(A_{23}^2-A_{22}A_{33}\Big)  \Big]^2, \label{d333}
\end{equation}
where $N$, is a non-zero function of the metric functions that cannot be written in compact form, and $D$ is proportional to the determinant of the metric\footnote{For $g^I$ and $g^{II}$, the proportionality constant is $2/\beta_\Phi^5\Phi^3$.} (\ref{metricaux}). Inserting the metric functions from Eqs.(\ref{g11})-(\ref{g33}), we obtain\footnote{We fixed the value of $k$ equal to zero.}

\begin{align}
    D^I&= 2 \beta_\Phi \Phi^3 \Big \{ \Phi_{,11}\Big [ \big (\Phi_{,23}\big)^2-\Phi_{,22}\Phi_{,33} \Big]+\Phi_{,22}\big (\Phi_{,13}\big)^2+\Phi_{,33}\big (\Phi_{,12}\big)^2-2 \Phi_{,12} \Phi_{,13} \Phi_{,23}   \Big\}^2, \label{d111}\\
     D^{II}&= 2 \beta_\Phi \Phi^3\big (\Phi_{,11}\big)^2\Big [ \big (\Phi_{,23}\big)^2-\Phi_{,22}\Phi_{,33} \Big]^2,\label{d222}\\
      D^{III}&=\Big \{  \Sigma_3U(\Phi,\partial\Phi)+\Sigma_1\Sigma_2V(\Phi,\partial\Phi) +\big(\Sigma_3\big)^2W(\Phi,\partial\Phi)  \Big\}^2. \label{d3}
\end{align}

For sake of simplicity, in $D^{III}$ we have defined the following functions 
\begin{equation}
    \Sigma_1=\beta_1 E^1 \Phi_{,1}\, \quad  \Sigma_2=\beta_2 E^2 \Phi_{,2}\quad,\quad \Sigma_3=\beta_3 E^3 \Phi_{,3}\quad,
\end{equation}
\bea 
    U\equiv  & -2\Sigma_1 \Sigma_2\Big\{\Phi_{,22}\big (\Phi_{,13}\big)^2+\Phi_{,11}\Big [(\Phi_{,23}\big)^2-2\Phi_{,22} \Phi_{,33}\Big]-\Phi_{,12}\Phi_{,13}\Phi_{,23}\Big\} \nonumber \\
    & + \Phi_{,12}\Big [ \big(\Sigma_1\big)^2+  \big(\Sigma_2\big)^2 \Big]\Big [\Phi_{,23}\Phi_{,13} -\Phi_{,12}\Phi_{,33}\Big],
\eea 
\be
      V \equiv  -\Sigma_1\Phi_{,22}\big(\Phi_{,13}\big)^2-\Phi_{,23}\Big[\Sigma_2\Phi_{,11}\Phi_{,23}-\Phi_{,12}\Phi_{,13}\Big(\Sigma_1+\Sigma_2\Big)\Big],
 \end{equation}
\begin{equation}
     W \equiv -\Sigma_1\Phi_{,11}\big(\Phi_{,23}\big )^2-\Phi_{,13}\Big[\Sigma_2\Phi_{,22}\Phi_{,13}-\Phi_{,12}\Phi_{,23}\Big(\Sigma_1+\Sigma_2\Big)\Big].
\end{equation}
Notice that, up to a conformal factor, $D^I$ and  $D^{II}$, corresponds to the square of the Hessian determinant of $\Phi$ computed for two different metrics. This result is not surprising, because it is known that in three dimensions the curvature of a metric is equivalent to the curvature of a Hessian metric \cite{pineda2019physical}, 
although it does not imply the existence of the underlying Hessian potential.

Notice that  $D^I$ and $D^{II}$ have the same structure as in the 
two-dimensional case. This allows us to extrapolate the above result for the curvature of the quasi-homogeneous metrics  $g^I$ and $g^{II}$ with and arbitrary number of   thermodynamic degrees of freedom, namely
\begin{align}
  R^I(E^1,E^2,E^3,\ldots, E^n)\quad &\propto  \quad\begin{vmatrix}
 \Phi_{,11} &  \Phi_{,12} & \ldots&  \Phi_{,1n}\\
\Phi_{,12} & \Phi_{,22} & \ldots &\Phi_{,2n}\\
\vdots &\vdots& \vdots&\vdots\\
\Phi_{,1n}&\Phi_{,2n}&\ldots&\Phi_{,nn}
\end{vmatrix}	^{-2},\\
R^{II}(E^1,E^2,E^3,\ldots, E^n)\quad &\propto  \quad \begin{vmatrix}
 \Phi_{,11} &  0 & \ldots&  0\\
0 & \Phi_{,22} & \ldots &\Phi_{,2n}\\
\vdots &\vdots& \vdots&\vdots\\
0&\Phi_{,2n}&\ldots&\Phi_{,nn}
\end{vmatrix}	^{-2}.
\end{align}
Nevertheless, for $R^{III}$ we could not find a neat expression. 

In the three-dimensional case under consideration, the scalar obtained from the square of the Ricci tensor\footnote{In a three-dimensional space, the Ricci tensor is the only relevant tensor since  the Weyl tensor vanishes identically.}, $\mathfrak{R}\equiv R_{ab}R^{ab}$, is an independent scalar and could have different curvature singularities. A direct
computation of $\mathfrak{R}$ , leads to
\begin{equation}
      \mathfrak{R}=\frac{H}{F},\quad  F\propto D^2,\label{d32}
\end{equation}
where $H$ again is a non-zero function of the metric variables that cannot be written in a compact form. Nevertheless, from the above equation, we observe that for the three metrics, the zeros of the denominator of $\mathfrak{R}$ are proportional to the ones of the scalar curvature, so they yield the same singularities. Moreover, for completeness, we computed the Kretschmann scalar, $\mathcal{K}_1=R_{abcd}R^{abcd}$, the Chern-Pontryagin scalar $\mathcal{K}_2=[^\star R_{abcd}]R^{abcd}$ and the Euler scalar $\mathcal{K}_3=[^\star R_{abcd}^\star] R^{abcd}$. It turns out that they are all proportional to the square of the scalar curvature. 

Therefore, we conclude that the singularities of the equilibrium space are determined by the zeros of the functions $D^I$, $D^{II}$ and $D^{III}$. We now analyze the zeros of these functions. The condition $D^{II}=0$, implies that $\big (\Phi_{,23}\big)^2=\Phi_{,22}\Phi_{,33}$ or $\Phi_{,11}=0$. We will refer to this as condition $II_A$ and $II_B$, respectively. For condition $II_A$ we have the following restrictions for $D^{III}$,
   \begin{align}
III_A&:\Phi_{,11}\Phi_{,22}\Phi_{,33}\Big\{ 2\Sigma_1\Sigma_2\Sigma_3-\Sigma_1\Big[(\Sigma_2\big)^2+(\Sigma_3\big)^2\Big]\Big\}\nonumber\\&+\Phi_{,12} \Phi_{,13} \Phi_{,23}
\Big\{(\Sigma_1\big)^2\Big[\Sigma_2+\Sigma_3\Big]+(\Sigma_2\big)^2\Big[\Sigma_1+\Sigma_3\Big]+(\Sigma_3\big)^2\Big[\Sigma_1+\Sigma_2\Big]+2\Sigma_1\Sigma_2\Sigma_3 \Big\}\nonumber\\
        &+\Phi_{,22}\big (\Phi_{,13}\big)^2\Big\{-\Sigma_2\Big[(\Sigma_1\big)^2+(\Sigma_3\big)^2\Big]-2\Sigma_1\Sigma_2\Sigma_3 \Big\}\\&+\Phi_{,33}\big (\Phi_{,12}\big)^2\Big\{-\Sigma_3\Big[(\Sigma_1\big)^2+(\Sigma_2\big)^2\Big]\Big\}=0\nonumber.
 \end{align}
 Notice that, if we demand $III_A=0$  for all values of quasi-homogeneous coefficients $\beta_a$, we must fulfill at least one of the following conditions
\begin{align}
&\Phi_{,22}=\Phi_{,33}=0 \implies \Phi_{,23}=0, \nonumber\\
&\Phi_{,11}=\Phi_{,12}=\Phi_{,13}=0,\nonumber\\
&\Phi_{,12}=\Phi_{,22}=0 \implies \Phi_{,23}=0,\label{c7}\\
&\Phi_{,13}=\Phi_{,33}=0 \implies \Phi_{,23}=0.\nonumber
\end{align}

Moreover, for condition $II_B$ ($\Phi_{,11}=0)$, the restrictions on $D^{III}$ read
        \begin{align}
III_B&:\Phi_{,12} \Phi_{,13} \Phi_{,23}
\Big\{(\Sigma_1\big)^2\Big[\Sigma_2+\Sigma_3\Big]+(\Sigma_2\big)^2\Big[\Sigma_1+\Sigma_3\Big]+ (\Sigma_3\big)^2\Big[\Sigma_1+\Sigma_2\Big]+2\Sigma_1\Sigma_2\Sigma_3\Big\}\nonumber\\
        &+\Phi_{,22}\big (\Phi_{,13}\big)^2\Big\{-\Sigma_2\Big[(\Sigma_1\big)^2+(\Sigma_3\big)^2\Big]-2\Sigma_1\Sigma_2\Sigma_3 \Big\}\nonumber\\&+\Phi_{,33}\big (\Phi_{,12}\big)^2\Big\{-\Sigma_3\Big[(\Sigma_1\big)^2+(\Sigma_2\big)^2\Big]\Big\}=0.
 \end{align}
If we demand $III_B$ to be zero for all the  values of the quasi-homogeneous coefficients $\beta_a$, we must fulfill at least one of the following conditions
\begin{align}
&\Phi_{,22}=\Phi_{,33}=0,\nonumber\\
&\Phi_{,12}=\Phi_{,13}=0, \label{kj}\\
&\Phi_{,12}=\Phi_{,22}=0,\nonumber\\
&\Phi_{,13}=\Phi_{,33}=0.\nonumber
\end{align}
We see that the singularities of the metrics  still are determined by the zeros of the second-order derivatives of $\Phi$. In fact, from Eqs.(\ref{c7}) and (\ref{kj}), we found a general set of conditions that relate all three singularities,  namely,
  \begin{align}
      I&:\Phi_{,11}\Big [ \big (\Phi_{,23}\big)^2-\Phi_{,22}\Phi_{,33} \Big]+\Phi_{,22}\big (\Phi_{,13}\big)^2+\Phi_{,33}\big (\Phi_{,12}\big)^2 -2 \Phi_{,12} \Phi_{,13} \Phi_{,23}=0,\label{c11}\\
       II&:\Phi_{,11}\Big [ \big (\Phi_{,23}\big)^2-\Phi_{,22}\Phi_{,33} \Big]=0,\label{vc}\\
       III&:\Phi_{,11}=\Phi_{,12}= \Phi_{,13}=0, \quad  \text{or}\nonumber \\
        & \ \  \Phi_{,12}= \Phi_{,22}=0, \quad  \text{or}\quad\Phi_{,13}= \Phi_{,33}=0, \quad \text{or}\quad\Phi_{,22}=\Phi_{,33}=0.\label{c33}\
 \end{align}
 
In general, for the metrics $g^I$ and $g^{II}$, we can express the denominator of the scalar curvature in a compact form using the Nambu bracket notation \cite{mansoori2015hessian}
\begin{equation}
    \begin{vmatrix}
 \Phi_{,11} &  \Phi_{,12} & \ldots&  \Phi_{,1n}\\
\Phi_{,12} & \Phi_{,22} & \ldots &\Phi_{,2n}\\
\vdots &\vdots& \vdots&\vdots\\
\Phi_{,1n}&\Phi_{,2n}&\ldots&\Phi_{,nn}
\end{vmatrix}=\Big\{\frac{\partial \Phi}{\partial E^1},\frac{\partial \Phi}{\partial E^2},\ldots ,\frac{\partial \Phi}{\partial E^n}\Big\}_{E^1,E^2\ldots E^n}
\end{equation}
where the Nambu bracket is defined as follows
\begin{align}
    \{f_1,f_2,\ldots, f_n\}_{E^1,E^2,\ldots, E^n}&=\sum^n_{ijk\ldots l=1}\epsilon_{ijk\ldots l}\frac{\partial f_1}{\partial E^i}\frac{\partial f_2}{\partial E^j}\frac{\partial f_3}{\partial E^k}\ldots \frac{\partial f_n}{\partial E^l},\quad \text{for}\quad n\geq 2,\\
    &=\frac{\partial f_1}{\partial E^1},\quad \text{for} \quad n=1.  
\end{align}
and $\epsilon_{ijk\ldots l}$ is the Levi-Civita symbol. For the two dimensional case, the above expression reduces to the Poisson bracket of two functions. Using this notation, the scalar curvature for the metrics $g^I$ and $g^{II}$ reads
\begin{align}
    R^I(E^1,E^2,\ldots, E^n)&\propto \Big\{\frac{\partial \Phi}{\partial E^1},\frac{\partial \Phi}{\partial E^2},\ldots ,\frac{\partial \Phi}{\partial E^n}\Big\}_{E^1,E^2\ldots E^n}^{-2},\\
     R^{II}(E^1,E^2,\ldots, E^n)&\propto \Phi_{,11}^{-2}\Big\{\frac{\partial \Phi}{\partial E^2},\ldots ,\frac{\partial \Phi}{\partial E^n}\Big\}_{E^2\ldots E^n}^{-2},
\end{align}
respectively. Thus, for  the metrics $g^I$ and $g^{II}$ the singularities of $R$ are identified with  the zeros of $\Big\{\frac{\partial \Phi}{\partial E^1},\frac{\partial \Phi}{\partial E^2},\ldots ,\frac{\partial \Phi}{\partial E^n}\Big\}$. Using the energy representation, i.e., $\Phi\equiv M$, and identifying $E^1\equiv S$. Then, we have that $\partial \Phi/\partial E^1\equiv T$, and  $I_a\equiv \partial \Phi/\partial E^a$, for $a=2,3,\ldots,n.$. Moreover, we can relate the singularities of the scalar curvature with the phase transition structure, as determined by the response functions,  of the  underlying thermodynamic system. In ordinary thermodynamics, the response functions define second-order phase transitions and are essentially determined by the behavior of the independent variables $E^a$ in terms of their duals $I_a$.
To doing that, we define the response functions \cite{mansoori2015hessian}
\begin{align}
C_{E_2,E_3,\ldots, E_n}&=T\Big(\frac{\partial S}{\partial T}\Big)_{E_2,E_3,\ldots, E_n}=T\frac{\{S,E_2,E_3,\ldots,E_n\}_{S,E^2,E^3,\ldots,E^n}}{\{T,E_2,E_3,\ldots,E_n\}_{S,E^2,E^3,\ldots,E^n}},\\
    C_{I_2,I_3,\ldots, I_n}&=T\Big(\frac{\partial S}{\partial T}\Big)_{I_2,I_3,\ldots, I_n}=T\frac{\{S,I_2,I_3,\ldots,I_n\}_{S,E^2,E^3,\ldots,E^n}}{\{T,I_2,I_3,\ldots,I_n\}_{S,E^2,E^3,\ldots,E^n}},\\
    \kappa_{S,I_3,\ldots,I_n}&=\Big(\frac{\partial E^2}{\partial I_2}\Big)_{S,I_3,\ldots,I_n}=\frac{\{S,E^2,I_3,\ldots,I_n\}_{S,E^2,E^3,\ldots,E^n}}{\{S,I_2,I_3,\ldots,I_n\}_{S,E^2,E^3,\ldots,E^n}},\\
    \kappa_{S,E^2,\ldots,I_n}&=\Big(\frac{\partial E^3}{\partial I_3}\Big)_{S,E^2,\ldots,I_n}=\frac{\{S,E^2,E^3,\ldots,I_n\}_{S,E^2,E^3,\ldots,E^n}}{\{S,E^2,I_3,\ldots,I_n\}_{S,E^2,E^3,\ldots,E^n}},\\ &\vdots\nonumber \\
    \kappa_{S,E^2,\ldots,E^n}&=\Big(\frac{\partial E^n}{\partial I_n}\Big)_{S,E^2,\ldots,E^n}=\frac{\{S,E^2,E^3,\ldots,E^n\}_{S,E^2,E^3,\ldots,E^n}}{\{S,E^2,E^3,\ldots,I_n\}_{S,E^2,E^3,\ldots,E^n}},\\
    \alpha_{S,E^2,\ldots E^n}&=\Big(\frac{\partial E^n }{\partial T}\Big)_{S,E^2,\ldots E^n}=\frac{\{S,E^2,\ldots,E^n\}_{S,E^2,E^3,\ldots,E^n}}{\{T,E^2,\ldots,E_n\}_{S,E^2,E^3,\ldots,E^n}}.
\end{align}
A direct calculation yields the following relation 
 \begin{align}
    R^I(E^1,E^2,\ldots, E^n)&\propto \Big[\frac{ C_{I_2,I_3,\ldots, I_n}\kappa_{S,I_3,\ldots,I_n} \kappa_{S,E^2,\ldots,I_n}\kappa_{S,E^2,\ldots,E^n}}{T}\Big]^2,\\
    R^{II}(E^1,E^2,\ldots, E^n)&\propto \Big[\frac{ C_{E_2,E_3,\ldots, E_n}}{T \kappa_{S,E^2,\ldots,I_n}\kappa_{S,E^2,\ldots,E^n\ldots I_n}}\Big]^2.
\end{align}
We do not have an expression that relates $R^{III}$ with the response functions of the thermodynamic system. However, we can express the condition $III$ using the response functions defined above. Thus, it is trivial to check that the singularities conditions $I$ (\ref{c11}), $II$ (\ref{vc}), and $III$ (\ref{c33}) in the energy representation take the following form
  \begin{align}
    I&: \frac{T}{C_{I_2,I_3}\kappa_{S,I_3}\kappa_{S,E^2}}=0 ,\label{re}\\
    II&: \frac{T\kappa_{S,I_3}\kappa_{S,E^2} }{C_{E_2,E_3}}=0,\\
    III&: \frac{1}{C_{E^2,E^3}}=\frac{1}{\alpha_{S,E^2}}=\frac{1}{\alpha_{S,E^3}}=0,\quad \text{or} \quad
    \frac{1}{\alpha_{S,E^3}}=\frac{1}{\kappa_{S,I_3}}=0,\label{re3}\\ \nonumber
    &\text{or} \quad \frac{1}{\alpha_{S,E^2}}=\frac{1}{\kappa_{S,E^2}}=0,\quad\text{or} \quad  \frac{1}{\kappa_{S,I_3}}=\frac{1}{\kappa_{S,E^2}}=0.
\end{align}

From the above results, we can conclude that in general the singularities of the equilibrium space are associated to the phase transitions of the response functions of the underlying thermodynamics system. 
Nevertheless, it is not clear how to classify them within the Ehrenfest scheme \cite{callen1998thermodynamics}, because the singularities of the equilibrium space cannot be associated to the divergence of a unique response function, rather to a specific combination of all of them. Thus, we might need a new scheme to classify the thermodynamic phase transitions in geometric/invariant manner.







\bibliographystyle{unsrt}

\end{document}